\documentclass{article}
\makeatletter

\usepackage{etoolbox}
\usepackage[T1]{fontenc}
\usepackage[margin=1in]{geometry}
\usepackage{placeins}
\usepackage{setspace}
\usepackage{amsmath}

\usepackage{graphicx}
\usepackage{hyperref}
\usepackage{url}
\usepackage{booktabs}
\usepackage{amsfonts}

\usepackage{microtype}
\usepackage{xcolor}
\usepackage[most]{tcolorbox}

\pdfmapline{+TencentSans-W7 < TencentSans-W7.pfb <T1-WGL4.enc}
\DeclareFontFamily{T1}{TencentSans-W7}{}
\DeclareFontShape{T1}{TencentSans-W7}{m}{n}{<-> TencentSans-W7}{}

\newcommand\tencentsanswseven[1]{{\usefont{T1}{TencentSans-W7}{m}{n} #1 }}

\definecolor{youtuBlue}{RGB}{0, 82, 217} \definecolor{youtuLightBlue}{RGB}{235, 245, 255} \definecolor{youtuText}{HTML}{1C2B33}
\hypersetup{
    colorlinks=true,
    linkcolor=youtuBlue,
    citecolor=youtuBlue,
    urlcolor=youtuBlue,
    pdfauthor={Tencent YouTu Lab},
    pdftitle={Tencent YouTu Lab Research Paper},
    pdfsubject={Artificial Intelligence},
    pdfkeywords={Tencent, YouTu Lab, AI, Research},
}

\renewcommand{\title}[1]{\newcommand{\titlelist}{\begin{center}{\LARGE\bf #1}\end{center}}}

\renewcommand\addtolist[5][]{
  \begingroup
    \if\relax#3\relax\def\sep{}\else\def\sep{#5}\fi
    \let\protect\@unexpandable@protect
    \xdef#3{\expandafter{#3}\sep #4[#1]{#2}}\endgroup
}

\newcommand\authorlist{}
\newcommand\authorformat[2][]{\begin{center}{{\tencentsanswseven{#2$^{#1}$}}}\end{center}}
\renewcommand\author[2][]{\addtolist[#1]{#2}{\authorlist}{\authorformat}{, }}

\newcommand\affiliationlist{}
\newcommand\affiliationformat[2][]{\begin{center}{\small{\tencentsanswseven{$^{#1}$#2}}}\end{center}}
\newcommand\affiliation[2][]{\addtolist[#1]{#2}{\affiliationlist}{\affiliationformat}{, }}

\newcommand\metadatalist{}
\newcommand\metadataformat[2][]{{\small {\bf #1:} #2}}
\newcommand\metadata[2][]{\addtolist[#1]{#2}{\metadatalist}{\metadataformat}{\par}}

\renewcommand\date[1]{\metadata[\raisebox{-0.1em}{\includegraphics[height=1em]{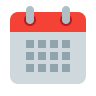}} Date]{#1}}
\newcommand\sourcecode[1]{\metadata[\raisebox{-0.1em}{\includegraphics[height=1em]{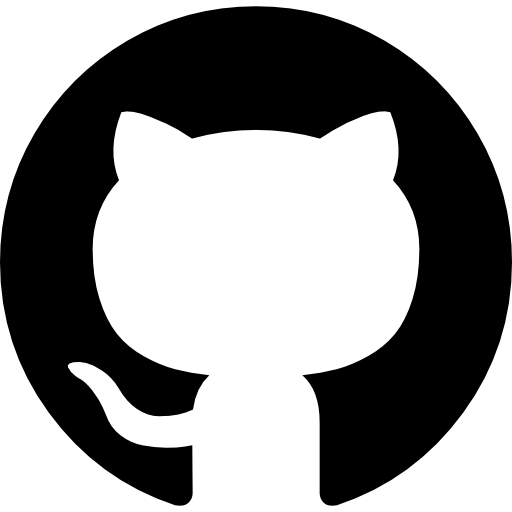}} Code]{\url{#1}}}

\newcommand\correspondence[1]{\metadata[\raisebox{-0.1em}{\includegraphics[height=1em]{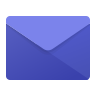}} Correspondence]{\texttt #1}}

\renewcommand{\abstract}[1]{\newcommand{\abstractlist}{{\color{youtuText} #1}}}

\newcommand{\mymaketitle}{\setlength{\parindent}{0cm}
  \setlength{\parskip}{0.5cm}
  {
    \setlength{\parskip}{0cm}
    \raggedright
    {
      \setstretch{1.618}
      \titlelist\par
    }
    \vskip 0.2cm
    \authorlist\par
    \vskip 0.2cm
    \affiliationlist\par

  }
  \begin{tcolorbox}[colback=youtuLightBlue, left=0.5cm, right=0.5cm, top=0.5cm, bottom=0.5cm, arc=10pt, enhanced, frame hidden, before skip=0pt, grow to left by=1.5pt, grow to right by=1.5pt]
    \abstractlist\par
    \vskip 0.5cm
    \ifdefempty{\metadatalist}{}{\metadatalist\par}
  \end{tcolorbox}
  \FloatBarrier
}

\renewcommand{\maketitle}{\mymaketitle}
\renewcommand{\section}{\@startsection{section}{1}{\z@}{-3.5ex \@plus -1ex \@minus -.2ex}{2.3ex \@plus.2ex}{\normalfont\Large\bfseries\color{youtuBlue}\tencentsanswseven}}
\renewcommand{\subsection}{\@startsection{subsection}{2}{\z@}{-3.25ex\@plus -1ex \@minus -.2ex}{1.5ex \@plus .2ex}{\normalfont\large\bfseries\color{youtuBlue}\tencentsanswseven}}
\renewcommand{\subsubsection}{\@startsection{subsubsection}{3}{\z@}{-3.25ex\@plus -1ex \@minus -.2ex}{1.5ex \@plus .2ex}{\normalfont\normalsize\bfseries\color{youtuBlue}\tencentsanswseven}}

\usepackage[font={small}]{caption}
\usepackage{fancyhdr}
\providecommand{\methodname}{}
\providecommand{\method}{}
\makeatother

\usepackage{mathpazo}
\usepackage{float}
\usepackage{needspace}
\usepackage[numbers,sort&compress]{natbib}
\usepackage{algorithm}
\usepackage{algpseudocode}
\usepackage{amsthm}
\usepackage{colortbl}
\definecolor{lmHeader}{RGB}{238,242,247}
\definecolor{lmOurs}{RGB}{232,243,253}
\definecolor{lmAccent}{RGB}{47,79,111}
\newtcolorbox{keytakeaway}{
    enhanced, breakable,
    colback=lmOurs, colframe=lmAccent, boxrule=0.6pt,
    left=4pt, right=4pt, top=3pt, bottom=3pt, arc=1.5pt,
    fontupper=\small,
}

\newtheorem{lemma}{Lemma}[section]
\renewcommand{\methodname}{LightningMoBA}
\renewcommand{\method}{\textbf{\methodname}}

\newcommand{\innergqa}{GQA-30B-A3B}

\newcommand{\papertitle}{Accelerating Mixture of Block Attention via Dynamic Routing, KV-Centric Execution and Runtime Pipelining}
\title{\fontsize{16}{19}\selectfont\papertitle}
\hypersetup{pdftitle={\papertitle}}
\author{Yingqi Shi, Junru Lu, Yun Liu, Di Yin, Xing Sun}
\affiliation{Tencent Youtu Lab}

\date{November 15, 2026}
\correspondence{junrulu@tencent.com}
\sourcecode{https://github.com/Tencent/LightningMoBA}

\begin{document}
\abstract{Mixture of Block Attention (MoBA) reduces long-context computation by selecting key-value (KV) blocks for each query.
FlashMoBA implements MoBA efficiently but retains a fixed Top-$k$ budget and query-centric schedule.
Fixed Top-$k$ applies one budget across different score distributions and couples routing to task granularity; query-range ownership fragments K/V reuse when several ranges select the same column.
We present \method{}, which co-designs selection and execution through dynamic routing, KV-centric ownership, and runtime pipelining.
Dynamic routing thresholds MoBA's per-query block scores and directly emits variable-cardinality sparse metadata.
KV-centric ownership groups selected rows by $(\text{query head},\text{KV block})$ column, splits high-fanout columns into bounded tasks, and reuses each task's loaded KV tile.
On Hopper, pipelining overlaps regular K/V transfers and irregular query gathers with Tensor Core computation.
Together, these mechanisms decouple retention from GPU scheduling granularity.
At matched Top-$16$ retention across five checkpoints, \method{} accelerates end-to-end generation by 1.20 to 1.60$\times$ while raising mean RULER score from 74.42 to 75.28.
}
\maketitle

\setlength{\parskip}{7pt}
\setlength{\intextsep}{10pt}
\setlength{\textfloatsep}{12pt}
\setlength{\floatsep}{10pt}
\captionsetup{skip=6pt}
\pretocmd{\section}{\par\Needspace{7\baselineskip}}{}{\PackageError{layout}{Section pagination patch failed}{}}
\pretocmd{\subsection}{\par\Needspace{6\baselineskip}}{}{\PackageError{layout}{Subsection pagination patch failed}{}}
\pretocmd{\subsubsection}{\par\Needspace{5\baselineskip}}{}{\PackageError{layout}{Subsubsection pagination patch failed}{}}
\pretocmd{\paragraph}{\par\Needspace{3\baselineskip}}{}{\PackageError{layout}{Paragraph pagination patch failed}{}}
\AtBeginEnvironment{lemma}{\Needspace{7\baselineskip}}
\AtBeginEnvironment{proof}{\Needspace{4\baselineskip}}

\vspace{-.1em}

\section{Introduction}
\label{sec:intro}
Dense self-attention evaluates all pairwise query-key interactions in a sequence, whereas sparse attention reduces this work by routing each query to only a subset of keys. On GPUs, however, fewer interactions do not automatically yield proportionally lower latency: score selection, sparse-index construction, irregular data movement, and load imbalance can offset the savings in arithmetic. An efficient sparse attention method must therefore jointly optimize the routing decision and the GPU workload that follows from it.

Mixture of Block Attention (MoBA) \citep{lu2025moba} routes at block granularity: it scores each KV block against the query, retains a fixed number $k$ of blocks, and computes softmax attention over their tokens. FlashMoBA \citep{xiao2025flashmoba} implements this efficiently, maintaining an ordered Top-$k$ candidate set per query and assigning each CTA a logical query range whose selected rows are gathered into dense Tensor Core tiles.

\begin{figure}[t]
    \centering
    \includegraphics[width=\linewidth]{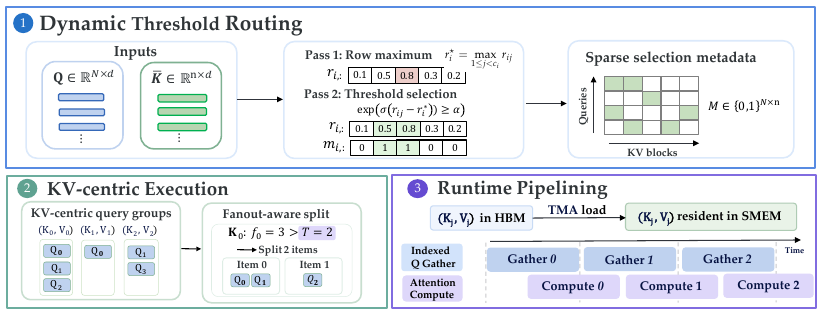}
    \caption{Overview of \method{}. Dynamic threshold routing adapts the retained
    block count per query and constructs sparse metadata directly; KV-centric
    execution organizes selected rows by $(\text{query head},\text{KV block})$
    column and splits high-fanout columns into bounded work items; runtime
    pipelining keeps each work unit's KV tile resident while its query gathers
    overlap Tensor Core work.}
    \label{fig:overview}
\end{figure}

Nevertheless, two limitations remain. Fixed Top-$k$ routing assigns the same block budget to queries with different score distributions, retaining weakly related blocks for concentrated scores while discarding relevant blocks for diffuse ones. Maintaining the ordered candidate set also adds routing latency. FlashMoBA organizes metadata by KV block but assigns CTA ownership by logical query range, so several CTAs may claim one $(\text{query head},\text{KV block})$ column and leave cross-CTA reuse to the cache hierarchy. Coupling query-group size to $k$ also makes a routing parameter determine scheduling granularity.

We present \method{}, which preserves MoBA's parameter-free query-block scoring rule and computes standard softmax attention over the retained tokens, but reorganizes routing and execution along three axes (Fig.~\ref{fig:overview}). Our contributions are as follows:
\begin{itemize}
    \item \textbf{Dynamic routing.} \method{} replaces the fixed per-query
    Top-$k$ budget with a threshold-based variable retention rule and emits
    sparse metadata during score evaluation, without an ordered candidate set
    or a learned auxiliary indexer.
    \item \textbf{KV-centric execution and runtime pipelining.} Work is organized
    by columns indexed by query head and KV block, so each unit reuses its
    loaded KV tile across the selected rows assigned to it. Fanout-aware
    partitioning bounds the rows per task, while an asymmetric Hopper pipeline
    overlaps TMA-based K/V transfers and irregular query gathers with Tensor
    Core computation.
    \item \textbf{Evaluation and analysis.} At matched Top-$16$ retention across
    five checkpoints, \method{} accelerates end-to-end generation by 1.20 to
    1.60$\times$ while increasing mean RULER score from 74.42 to 75.28.
    Controlled complete-device-path replays on two models reach 2.16 to
    2.56$\times$ speedup at Top-$32$, while routing and CSC construction reach
    6.99 to 11.25$\times$ at Top-$64$. Same-backend comparisons distinguish
    execution-backend gains, which dominate at Top-$16$, from dynamic routing's
    increasing benefit at larger budgets: routing alone yields 1.16 to
    1.56$\times$ end-to-end speedups at Top-$64$ on the evaluated two models
    (Sec.~\ref{sec:exp:ablation:routing-body}). Training integration using the
    adapted FlashMoBA backward implementation achieves a 1.602$\times$ speedup
    on Qwen3-4B at Top-$64$ with closely aligned observed training losses
    (Sec.~\ref{sec:exp:training-performance}).
\end{itemize}

\section{Related Work}
\label{sec:related}
The efficiency of sparse attention is determined jointly by its selection rule
and by the execution path that consumes the resulting sparse pattern.

\subsection{Scoring, Granularity, and Budgeting in Sparse Attention}
\label{sec:related:taxonomy}

Prescribed patterns \citep{beltagy2020longformer,zaheer2020bigbird} fix
connectivity by design; input-derived patterns
\citep{kitaev2020reformer,roy2021routing} cluster or hash tokens, and SSA
\citep{shen2025ssa} aligns sparse- with full-attention representations during
training. Within content-dependent selection, granularity, scoring, and budget
are separate choices: MoBA \citep{lu2025moba} scores blocks with mean-pooled
keys under a fixed per-query budget; NSA \citep{yuan2025nsa} combines
compressed, selected, and sliding-window branches; DSA \citep{deepseekai2025dsa}
and MSA \citep{lai2026msa} use learned indexers, the latter per GQA group;
HiLS \citep{hu2026hils}, HySparse \citep{gao2026hysparse} and HySparse2
\citep{wei2026hysparse2} learn summaries or derive routing from a preceding
full-attention layer. Candidate discovery is amortized along different axes
\citep{qiu2026qsa,liu2026pivot,bai2026indexcache,deepseekai2026csa2}, and
adaptive budgets also appear \citep{yuan2025blasst,wu2026vmoba}. Thresholding
avoids sorting: FlashPrefill \citep{fan2026flashprefill} shares one
block-level cutoff per query block, while \method{} thresholds MoBA's
per-query block scores, so retention differs across individual queries.
FlexPrefill \citep{lai2025flexprefill} adapts each head's sparse pattern
to the input context, switching between query-specific and predefined patterns.
Its cumulative-attention-based index selection retains query--key indices until
a target attention mass is reached. In contrast, \method{} preserves MoBA's
block scorer and uses per-query block-score thresholds with KV-centric execution.
Retrofitting pretrained checkpoints through a learned selector is another route
\citep{zhou2026rtpurbo,ahmad2026spotattention}. \method{} changes neither
weights nor scorer; it replaces only the rule that maps scores to a retained set
and co-designs the representation and execution of the resulting variable
pattern.

\subsection{Co-Design of Sparse Attention and Systems}
\label{sec:related:systems}

The cost of a pattern depends on how its interactions are represented and scheduled: FlashMoBA \citep{xiao2025flashmoba} assigns CTA ownership by logical query range; FAST-Prefill \citep{jayanth2026fastprefill} co-designs index generation and caching for an FPGA. KV-outer execution is established \citep{lu2025moba,yan2025fsa,lai2026msa}: fixed Top-$k$ fixes the query-side row degree but leaves the KV-side column degree varying, so reverse indices and splitting popular columns are existing techniques.

\method{} extends KV-outer execution to routing graphs with variable
query-side degree: under per-query thresholding, both the row degree of each
query and the total number of retained pairs are set at run time. It co-designs the routing representation and execution interface, decoupling the
per-query budget from the kernel's logical task size and dividing high-fanout
columns into tasks with bounded query rows.

\subsection{High-Performance Kernels on Modern GPUs}
\label{sec:related:kernels}

FlashAttention \citep{dao2022flashattention,dao2023flashattention2} tiles attention with online softmax; FlashAttention-3 \citep{shah2024flashattention3} and FlashAttention-4 \citep{zadouri2026flashattention4} add warp-specialized producer-consumer pipelines on Hopper and Blackwell. PagedAttention \citep{kwon2023pagedattention} manages KV-cache
storage at the serving level, orthogonal to the organization of data movement
inside one attention call. These kernels show that tile shape, work ownership, and operand access determine what can overlap. \method{} applies these principles to a KV-centric sparse
workload with asymmetric operand access: contiguous K/V tiles move through TMA
while indexed query rows move through a dedicated producer path, both
overlapping Tensor Core computation, with tile shapes and warp roles independent of how many blocks each query retains.

\section{Methodology}
\label{sec:method}

\subsection{Preliminaries: MoBA and FlashMoBA}
\label{sec:method:prelim}

We consider causal prefill over $N$ tokens with head dimension $d$. Batch and
head indices are omitted unless needed, and $g$ query heads share each key/value
head under GQA. We write $\mathbf{q}_i$, $\mathbf{k}_t$ and $\mathbf{v}_t$ for a token's
query, key and value vectors, reserving the unsubscripted $k$ for the number of
blocks a query retains, with bold type keeping the two apart.

Mixture of Block Attention (MoBA) \citep{lu2025moba} partitions the sequence into
$n=\lceil N/B\rceil$ blocks of at most $B$ tokens; the keys of block $j$ form
$\mathbf{K}_j\in\mathbb{R}^{B\times d}$. For each query $\mathbf{q}_i$, MoBA
scores each block against the query through the block's mean-pooled key and
keeps the best $k$ causally available blocks, including the query's own block
$c_i=\lceil i/B\rceil$:
\begin{equation}
    \tilde{\mathbf{k}}_j = \frac{1}{|\mathbf{K}_j|}\sum_{\mathbf{k}\in\mathbf{K}_j}\mathbf{k},
    \qquad
    r_{ij} = \mathbf{q}_i^\top \tilde{\mathbf{k}}_j,
    \qquad
    \mathcal{S}_i = \{c_i\}\cup
    \operatorname*{TopK}_{\min\{k-1,\,c_i-1\}}
    \bigl(\{r_{ij}\}_{1\le j<c_i}\bigr).
    \label{eq:moba-selection}
\end{equation}
Pooling is used only for routing; attention is computed over all tokens in the
retained blocks. We write
$Z=\sum_i|\mathcal{S}_i|$
for the number of selected query-block pairs, i.e., the sparse pattern executed
by the attention stage. Under Eq.~\eqref{eq:moba-selection} each query retains
$1+\min\{k-1,\,c_i-1\}\le k$ blocks, so the budget bounds the pattern size as
well: $Z\le Nk$. Sec.~\ref{sec:method:routing} replaces the fixed budget by a
score criterion, after which $|\mathcal{S}_i|$ and $Z$ are set by the scores
rather than by $k$. Sec.~\ref{sec:method:complexity} gives the corresponding worst-case bound.

FlashMoBA \citep{xiao2025flashmoba} provides an efficient GPU implementation
of MoBA. It evaluates the fixed Top-$k$ routing rule, constructs the
corresponding sparse metadata, and executes block-sparse attention over the
selected query-block pairs. Its reference router maintains a running Top-$k$
candidate set for each query, while the attention kernel organizes execution
around logical query ranges and gathers selected query rows into dense Tensor
Core tiles.

We retain MoBA's query-block score based on mean-pooled keys and its standard
softmax attention over retained tokens, while redesigning block selection and
the sparse workload that follows.

\subsection{Dynamic Routing}
\label{sec:method:routing}

MoBA assigns the same routing budget $k$ to every query, regardless of how its
block scores are distributed. This fixed-cardinality constraint need not match
the amount of relevant context available to each query: for a concentrated
score distribution, selecting exactly $k$ blocks may retain weakly related
blocks only to satisfy the budget, whereas for a more diffuse distribution the
same budget may discard additional blocks with non-negligible scores.
Top-$k$ therefore fixes how many blocks are retained rather than what score
level is sufficient for retention.

\begin{figure}[t]
    \centering
    \includegraphics[width=\linewidth]{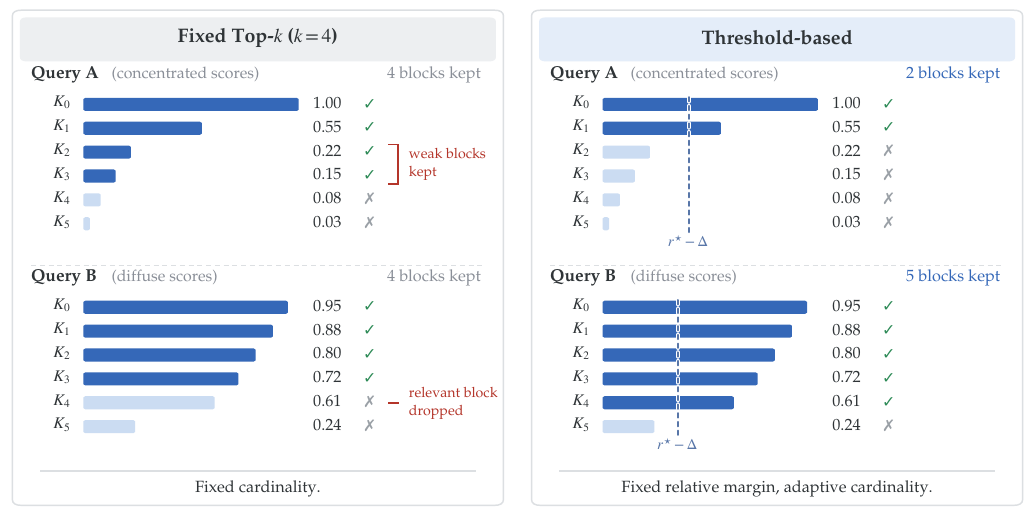}
    \caption{Fixed Top-$k$ and threshold-based selection applied to the same
    block scores of two queries. A fixed budget keeps the same number of blocks
    for both, retaining weakly scored blocks when the scores are concentrated
    and discarding a comparably scored one when they are diffuse. The threshold
    rule keeps every block within a fixed margin
    $\Delta=-\log\alpha/\sigma$ of each query's strongest preceding-block score, so
    the retained count follows the score distribution. Compulsory blocks are omitted so that the comparison isolates the selection rule.}
    \label{fig:selection-rule}
\end{figure}

Inspired by FlashPrefill \citep{fan2026flashprefill}, we instead apply a
query-local threshold to MoBA's original block scores. The scoring function is
unchanged; only the rule that converts scores into a retained set is replaced.
As illustrated in Fig.~\ref{fig:selection-rule}, this fixes a relative score
criterion while allowing the retained cardinality to adapt independently for
each query.

\subsubsection{Threshold-Based Block Selection}
\label{sec:method:threshold}

Query $\mathbf{q}_i$ may attend to the causally admissible blocks
$\mathcal{C}_i=\{j:1\le j\le c_i\}$. We retain unconditionally up to
$n_{\mathrm{sink}}$ initial sink blocks and a local window of $w\ge1$ blocks ending
at the query's own block,
\begin{equation}
    \mathcal{F}_i
    =
    \{1,\ldots,\min(n_{\mathrm{sink}},c_i)\}
    \cup
    \{\max(1,c_i-w+1),\ldots,c_i\}
    \subseteq
    \mathcal{C}_i .
    \label{eq:forced-blocks}
\end{equation}
Let $\sigma>0$ denote the softmax scale and
$r_i^{\star}=\max_{1\le j<c_i} r_{ij}$ the largest score over complete
blocks preceding the query's own block. If $c_i=1$, set $r_i^{\star}=0$;
there are no optional blocks. We retain
\begin{equation}
    \mathcal{S}_i
    =
    \mathcal{F}_i
    \cup
    \Bigl\{
        j\in\{1,\ldots,c_i-1\}\setminus\mathcal{F}_i :
        \exp\bigl(\sigma(r_{ij}-r_i^{\star})\bigr)\ge\alpha
    \Bigr\},
    \qquad
    \alpha\in(0,1].
    \label{eq:threshold-rule}
\end{equation}

Equivalently, taking logarithms on both sides of the inequality in
Eq.~\eqref{eq:threshold-rule} and dividing by $\sigma$, an optional block is
retained when
\begin{equation}
    r_i^{\star}-r_{ij}
    \le
    -\frac{\log\alpha}{\sigma}
    =:
    \Delta ,
\end{equation}
i.e., when its score lies within a fixed margin $\Delta$ of the strongest
preceding block for that query. The decision therefore requires \textbf{no softmax
normalization}. Unlike fixed-budget routing, which fixes cardinality and induces
a query-dependent score cutoff, our rule fixes a query-relative score criterion
and lets $|\mathcal{S}_i|$ vary across queries and heads.

The current block is retained with token-level causal masking, but its pooled
key is excluded from the reference maximum because it can contain future
tokens. Preceding forced blocks still contribute to the maximum. The rule
depends only on score differences relative to $r_i^{\star}$, so it is invariant
to an additive shift of the preceding-block scores for a query.

The router emits this representation with a two-pass streaming construction and a prefix-sum scatter into compact CSC metadata; we defer those details to Appendix~\ref{sec:router-details} and only use their output here.

\subsection{KV-Centric Execution}
\label{sec:method:execution}

The sparse pattern produced by routing is content-dependent, whether or not its per-query cardinality is fixed. FlashMoBA assigns work by logical query ranges, so queries in different ranges that select the same KV block are handled by different CTAs, causing repeated KV loads and fragmented reuse across query-owned tasks. This limits the useful work each loaded KV tile can amortize.

We instead organize the sparse pattern by $(\text{query head},\text{KV block})$
columns and assign execution around these columns
(Sec.~\ref{sec:related:systems}). Queries selecting the same block under a query head are therefore grouped under
the same KV-owned column, allowing each work unit to reuse its loaded KV tile
across its assigned rows. Because a column's task size is proportional to its
fanout, assigning an entire column to one CTA can leave a long-running tail. We
therefore separate two decisions: \emph{ownership}, which determines which KV
column a task belongs to, and \emph{task size}, which partitions high-fanout
columns into bounded work units.
Alg.~\ref{alg:kvcentric} illustrates the KV-centric ownership pattern before
fanout-aware splitting: execution is organized by KV columns rather than query
ranges. Sec.~\ref{sec:method:fanout} then partitions high-fanout columns into
bounded work items without changing this ownership rule.

\subsubsection{KV-Centric Ownership}
\label{sec:method:kvcentric}

In FlashMoBA a CTA, scheduled as a unit onto an SM, owns a band of $M$
consecutive query rows and walks the blocks those rows selected, gathering only
that band's rows so that routing induces no masked lanes. Write
\begin{equation}
    \mathcal{Q}_{h,j}=\{\,i : j\in\mathcal{S}_{h,i}\,\},
    \qquad
    \lambda_{h,j}(M)=\bigl|\{\,b : \mathcal{Q}_{h,j}\cap[bM,(b{+}1)M)\neq\emptyset\,\}\bigr|,
    \label{eq:ownership}
\end{equation}
for the queries that selected column $(h,j)$ and the number of bands claiming
it, where a column is block $\mathcal{B}_j$ as seen by query head $h$. With $C$
selected columns, the mean over columns
\begin{equation}
    \Lambda(M)=\frac{1}{C}\sum_{h,j}\lambda_{h,j}(M)
    \;=\;\frac{|\mathcal{V}(M)|}{C},
    \qquad
    \mu(M)=\frac{Z}{|\mathcal{V}(M)|}=\frac{\bar{f}}{\Lambda(M)},
    \qquad
    \bar{f}=\frac{Z}{C},
    \label{eq:lambda}
\end{equation}
counts how many times a column is claimed, where $\mathcal{V}(M)$ is the set of
(band, column) pairs carrying work and $\bar{f}$ is the mean fanout. These two
quantities determine the costs of band ownership. A column's K/V is requested
$\lambda_{h,j}(M)$ times rather than once. Since a physical KV block may appear
in up to $g$ query-head columns, band fragmentation compounds with GQA sharing:
$\Lambda(M)$ measures the mean multiplicity at the column level, while the
request count of a physical block is the sum of the multiplicities of its
participating columns, and how many of those requests reach HBM depends on cache
behaviour. Independently, each request exposes only $\mu(M)$ selected rows with which to fill an MMA tile that holds $m_{\mathrm{MMA}}$ rows. The two are not
independent: $Z$ and $C$ are fixed by the routing pattern, and $M$ changes only
$|\mathcal{V}(M)|$, so dividing numerator and denominator by $C$ gives
\[
    \mu(M)
    = \frac{Z}{|\mathcal{V}(M)|}
    = \frac{Z/C}{|\mathcal{V}(M)|/C}
    = \frac{\bar{f}}{\Lambda(M)} .
\]
At fixed $\bar{f}$, raising $\Lambda$ multiplies the requests per column and
divides the work each request carries by the same factor. The arithmetic is
unchanged: fragmenting leaves the routing pattern untouched, so the same $Z$
query-block pairs must still be evaluated, and $C$, $\bar{f}$ and the
$\mathcal{O}(ZBd)$ term of Sec.~\ref{sec:method:complexity} are unaffected.
What changes is how often the K/V operand is fetched and how much of each MMA tile is filled with work that is actually useful.

Within fixed query-band ownership, increasing $M$ is the direct way to reduce this fragmentation (shared memory caps it, but the cap is not what makes $M$ the wrong knob): a band claims a column alone only if it contains every query that selected it. The resulting necessary condition is stated below, with
$s_{h,j}=\max\mathcal{Q}_{h,j}-\min\mathcal{Q}_{h,j}+1$ denoting the query span
of a column.
\begin{lemma}[Span bound for unit multiplicity]\label{lem:span}
Unit multiplicity requires
\begin{equation}
    \Lambda(M)=1 \quad\Longrightarrow\quad M\ \ge\ \max_{h,j} s_{h,j}.
    \label{eq:span-bound}
\end{equation}
The converse does not hold.
\end{lemma}
The requirement is not an artifact of how $M$ is chosen. Band ownership groups
queries by position, whereas routing groups them by content, so the two
partitions coincide only by accident; a column selected by both an early and a
late query spans the sequence, and no band shorter than that span can own it
alone. This is a property of content-based routing rather than of thresholding, so it
also applies to a fixed budget. It explains why query-band ownership works for
local or structured sparsity, where each block's consumers are contiguous by
construction, but fragments content-dependent patterns.
Fig.~\ref{fig:ownership} evaluates Eq.~\eqref{eq:lambda} on captured routing
patterns; we use it to quantify how much fragmentation remains at practical values of $M$.

Making the selected column $(h,j)$ the ownership domain removes band
fragmentation: all selected rows of a column are grouped under the same KV owner
before tasks are formed, and $M$ leaves the ownership decision entirely. This
gives unit multiplicity at the column level before deliberate fanout splitting,
$\lambda_{h,j}=1$, and makes every selected row available for tile formation.
It does not fix how many CTAs serve the column: Sec.~\ref{sec:method:fanout}
partitions it into several work units that each load the tile for their assigned
rows, emit a partial output and local log-sum-exp, and are combined by the
online-softmax merge, which is exact in real arithmetic.

\subsubsection{Fanout-Aware Workload Decomposition}
\label{sec:method:fanout}

KV-centric ownership removes band fragmentation by grouping all consumers of a
column under one ownership domain. A column's work grows with its fanout
$f_{h,j}=|\mathcal{Q}_{h,j}|$, the number of queries selecting it. A task
covering the column serially processes
\begin{equation}
    w_{h,j}=\bigl\lceil f_{h,j}/m_{\mathrm{MMA}}\bigr\rceil
    \label{eq:worksize}
\end{equation}
MMA tiles, so fanout becomes task size and dispersion in fanout becomes
dispersion in task duration. Fixed Top-$k$ bounds how many blocks a query keeps,
not how many queries keep a block, so this dispersion is a property of the
routing rather than of thresholding.

Dispersion in task size affects scheduling differently from total work.
Lemma~\ref{lem:makespan} in Appendix~\ref{sec:method-details} specializes the
classical list-scheduling bound to these work units: Eq.~\eqref{eq:makespan}
separates an aggregate throughput term $W/P$ from a largest-work-item term
$\max_{h,j} w_{h,j}$. Fanout dispersion raises the latter. Additional hardware
parallelism reduces the first term but cannot subdivide the largest item, and
tensor parallelism sharpens this tail because a synchronized step waits for its
slowest device.

Fig.~\ref{fig:fanout-halfcol} characterizes the source of that tail on captured
routing patterns. It reports the mean and selected column-fanout quantiles for
one high-tail prefill trace, normalized by its mean of 2,031 query rows. The
median is only $0.064\times$ the mean, whereas P99 and the maximum reach
$28.1\times$ and $64.5\times$ the mean. This dispersion raises the
largest-work-item term in Eq.~\eqref{eq:makespan}. The corresponding control
trade-off is shown in Fig.~\ref{fig:fanout} in Appendix~\ref{sec:add-figures}:
decreasing the task-size cap $T$ reduces the largest work item while increasing
load multiplicity. Both figures use the same actual all-pair columns, including
compulsory sink and local selections; they characterize execution work rather
than content-selected fanout alone. The optional-only tail is reported separately
in Appendix~\ref{sec:exp:ablation:fanout-diagnostics}.

\paragraph{Fanout-aware splitting} To prevent a high-fanout column from defining
the critical path, we decouple ownership from task size. Ownership remains
KV-centric; only the granularity of individual work units changes. Capping the
number of selected query rows a unit may claim, we partition each column into
work units $(h,j,i_{\text{begin}},i_{\text{count}})$ subject to
$i_{\text{count}}\le T$.
The cap is on the count of selected rows in the column's compact row list, not on
the sequence positions those rows occupy, so it is unrelated to the span
$s_{h,j}$ of Sec.~\ref{sec:method:kvcentric}. A count cap suffices because the
metadata of Sec.~\ref{sec:method:nnz} lists a column's selected rows contiguously
even when those queries are scattered over the input sequence.

The exchange is explicit in both directions: $\max_{h,j} w_{h,j}$ is bounded by
$\lceil T/m_{\mathrm{MMA}}\rceil$, while a column of fanout $f_{h,j}$
becomes $\lceil f_{h,j}/T\rceil$ units that each independently load the column's
K/V operand and produce the outputs for their assigned query rows, raising the
multiplicity to
\begin{equation}
    \Lambda_T=\frac{1}{C}\sum_{h,j}\bigl\lceil f_{h,j}/T\bigr\rceil
    \;<\; 1+\bar{f}/T .
    \label{eq:lambda-t}
\end{equation}
Fanout-aware splitting therefore trades within-unit KV reuse for finer-grained
parallelism and improved load distribution, rather than obtaining them alongside
it, and both sides of the trade are computable from the routing metadata before
the attention work is launched: lowering $T$ reduces the largest-task term in
Eq.~\eqref{eq:makespan} while increasing $\Lambda_T$.
Taking $T$ to be a multiple of $m_{\mathrm{MMA}}$ preserves the total number of
MMA row tiles for each column while redistributing those tiles among
independently schedulable work units.
Sec.~\ref{sec:exp:ablation:fanout} (Table~\ref{tab:fanout-split-ablation})
sweeps $T$ and reports the setting we use.

\subsection{Runtime Pipelining}
\label{sec:method:scheduling}

KV-centric execution creates asymmetric operand access: K/V tiles remain
contiguous, whereas Q rows must be gathered from the routing metadata. We assign
regular K/V transfers to the Tensor Memory Accelerator (TMA) and irregular Q
gathers to a producer path, allowing both paths to be staged concurrently with
Tensor Core computation. The implementation targets Hopper's TMA, WGMMA and
warp-specialized execution model; Sec.~\ref{sec:discussion} discusses
portability.

\section{Experiments}
\label{sec:exp}

\subsection{Experimental Setup}
\label{sec:exp:setup}

We evaluate end-to-end inference and training performance, controlled execution and routing costs, and long-context quality. Inference comparisons are training-free; training integration is evaluated separately under the settings in Sec.~\ref{sec:exp:training-performance}. Model configurations, benchmark settings, and timing scopes are specified below and in the corresponding result sections.

\paragraph{Models}
The evaluation uses five checkpoints: Qwen3-4B-Instruct-2507 (Qwen3-4B),
Qwen3-235B-A22B, Qwen3-32B-YaRN4,
Llama-3.1-8B-Instruct (Llama-3.1-8B), and \innergqa{}, an internal GQA model
(30B total, 3B active) whose architecture follows Qwen3 and is trained from
scratch. These span dense and mixture-of-experts architectures as well as
different GQA grouping ratios; checkpoint sources are listed with the full
setup in Appendix~\ref{sec:setup-details}.
Each reported configuration identifies the checkpoint revision, tested
context lengths, head configuration, and the layers to which sparse attention is applied.

We compare against an optimized implementation of the same block-attention method; its baseline definition, training-free protocol and model sources are in Appendix~\ref{sec:setup-details}.

\Needspace{100pt}
\subsection{End-to-End Performance}
\label{sec:exp:main}

\subsubsection{End-to-End Inference Performance}
\label{sec:exp:main:speed}

Table~\ref{tab:main-e2e} reports the end-to-end comparison on RULER at the
Top-$16$ reference budget.
The baseline is FlashMoBA with fixed Top-$16$ routing, one sink block and a
two-block local window; \method{} uses dynamic threshold prefill with the same
compulsory blocks and a matched average Top-$16$ retention budget.
Both variants use dense decoding with CUDA Graph enabled.
Every entry aggregates the full 6,500-example RULER suite
at tensor parallelism eight and batch size one, and times are generation-loop
wall times from the first request to the completion of the last, excluding
model loading, data preparation and result serialization.
Top-$16$ is the routing budget used by the deployed baseline.
We further extend the end-to-end evaluation to the Top-$32$ reference budget using
the same five checkpoints and full 6,500-example RULER suite
(Appendix~\ref{sec:exp:top32}, Table~\ref{tab:main-e2e-top32}).
\methodname{} achieves lower generation time in all 30 model--length configurations,
with per-model average speedups of 1.23--1.49$\times$ over FlashMoBA.
Controlled routing-cost measurements across different retention budgets are
reported separately in Section~\ref{sec:exp:ablation:routing-body}.

\suppressfloats[t]
\begin{table}[!htbp]
    \centering
    \caption{End-to-end RULER results at the Top-$16$ reference budget over
    the full 6,500-example suite.
    Each cell reports Base / Ours; wall times are in minutes with speedup in
    parentheses. Averages are over the six context lengths for each model,
    and the better value is bolded.}
    \label{tab:main-e2e}
    \small
    \setlength{\tabcolsep}{2.6pt}
        \resizebox{\linewidth}{!}{\begin{tabular}{lccccc}
        \toprule
        \rowcolor{lmHeader}
        Length & Qwen3-4B & \innergqa{} & Qwen3-235B-A22B
            & Llama-3.1-8B & Qwen3-32B-YaRN4 \\
        \midrule
        \multicolumn{6}{c}{\textit{(a) RULER score: Base / Ours}} \\
        \midrule
        4K & 93.75 / 94.29 & 94.40 / 94.10 & 95.26 / 94.86 & 95.27 / 94.96 & 89.15 / 88.98 \\
        8K & 85.98 / 91.87 & 88.20 / 89.15 & 94.27 / 94.65 & 91.82 / 93.54 & 85.90 / 87.80 \\
        16K & 79.45 / 87.18 & 82.47 / 80.88 & 89.18 / 93.26 & 83.62 / 87.25 & 74.97 / 74.11 \\
        32K & 69.98 / 79.93 & 74.09 / 70.82 & 79.72 / 84.03 & 70.73 / 77.76 & 67.44 / 69.96 \\
        64K & 59.08 / 53.41 & 57.14 / 55.72 & 70.87 / 68.78 & 61.05 / 57.57 & 61.58 / 56.32 \\
        128K & 46.35 / 46.51 & 34.92 / 35.40 & 63.78 / 62.90 & 41.45 / 41.75 & 50.85 / 50.54 \\
        \midrule
        \emph{Average} & 72.43 / \textbf{75.53} & \textbf{71.87} / 71.01 & 82.18 / \textbf{83.08} & 73.99 / \textbf{75.47} & \textbf{71.65} / 71.29 \\
        \midrule
        \multicolumn{6}{c}{\textit{(b) Wall time: Base / Ours (speedup)}} \\
        \midrule
        4K & 145.9 / 60.5 (2.41$\times$) & 119.1 / 36.8 (3.24$\times$) & 325.6 / 123.9 (2.63$\times$) & 135.6 / 70.9 (1.91$\times$) & 270.6 / 171.6 (1.58$\times$) \\
        8K & 153.1 / 69.5 (2.20$\times$) & 118.7 / 41.1 (2.89$\times$) & 368.6 / 150.8 (2.44$\times$) & 142.7 / 75.6 (1.89$\times$) & 290.3 / 195.0 (1.49$\times$) \\
        16K & 165.0 / 87.7 (1.88$\times$) & 116.5 / 59.3 (1.96$\times$) & 458.2 / 206.0 (2.22$\times$) & 150.7 / 88.2 (1.71$\times$) & 317.8 / 240.0 (1.32$\times$) \\
        32K & 174.8 / 114.3 (1.53$\times$) & 146.2 / 103.9 (1.41$\times$) & 579.3 / 341.2 (1.70$\times$) & 166.1 / 117.0 (1.42$\times$) & 394.1 / 327.7 (1.20$\times$) \\
        64K & 206.7 / 153.1 (1.35$\times$) & 178.8 / 122.5 (1.46$\times$) & 785.0 / 537.1 (1.46$\times$) & 192.8 / 158.3 (1.22$\times$) & 504.1 / 446.8 (1.13$\times$) \\
        128K & 284.2 / 251.6 (1.13$\times$) & 293.0 / 253.1 (1.16$\times$) & 1180.1 / 957.4 (1.23$\times$) & 277.9 / 241.3 (1.15$\times$) & 812.3 / 770.9 (1.05$\times$) \\
        \midrule
        \emph{Average} & 188.3 / \textbf{122.8} (\textbf{1.53}$\times$) & 162.1 / \textbf{102.8} (\textbf{1.58}$\times$) & 616.1 / \textbf{386.1} (\textbf{1.60}$\times$) & 177.6 / \textbf{125.2} (\textbf{1.42}$\times$) & 431.5 / \textbf{358.7} (\textbf{1.20}$\times$) \\
        \bottomrule
    \end{tabular}}
\end{table}
\begin{keytakeaway}
\noindent\textbf{Key result.}
At the Top-$16$ budget used by the deployed baseline, \method{} shortens the
total generation-loop time of all five models, with per-model Average speedups
of 1.20 to 1.60$\times$ (Table~\ref{tab:main-e2e}). Mean RULER score rises for
three models and falls for two, by 0.86 points on \innergqa{} and 0.36 points
on Qwen3-32B-YaRN4. Quality also varies with length: the largest gains occur
between 8K and 32K, the two methods stay within one point of each other at 4K
and 128K, and at 64K \method{} scores 1.41 to 5.67 points lower on all five
models.
\end{keytakeaway}

Additional comparisons with MoBA and FlexPrefill are provided in
Appendix~\ref{sec:sparse-baselines}.

\FloatBarrier
\subsubsection{End-to-End Training Performance}
\label{sec:exp:training-performance}

\paragraph{Training integration.}
We further extend \methodname{} to end-to-end training and evaluate its performance
on Qwen3-4B. The forward pass uses dynamic threshold routing and the optimized
KV-centric attention implementation, while the backward pass adapts the original
FlashMoBA CUDA kernels to consume dynamically generated CSC metadata and merged
forward outputs, including the global log-sum-exp (LSE). The backward implementation
also handles gradient accumulation and grouped key/value gradient reductions for
GQA. Routing decisions are treated as non-differentiable, and gradients are
computed over the retained sparse attention pattern. We compare dynamic threshold
routing against fixed TopK under three matched average retention budgets: Top-16,
Top-32, and Top-64. Both configurations share the optimized KV-centric forward
and adapted backward implementation; only the routing strategy differs.

\paragraph{Training setup.}
All six runs start from the same Qwen3-4B-Instruct-2507 weights and use identical
PG19 windows at 128K for 100 training steps (seed 1234). We use tensor parallelism
of 1, global and micro batch sizes of 1, BF16 precision, and AdamW with a learning
rate of $10^{-5}$ and weight decay of 0.1. Block size is 128; one sink block and a
two-block local window count toward the retention budget. Thresholds are
calibrated on disjoint PG19 windows and fixed throughout each run. Reported
averages include all 100 steps. TFLOPS/GPU is an effective throughput metric:
analytically estimated sparse GEMM FLOPs, including model linear layers and
retained attention, are divided by the measured end-to-end training step time and
GPU count, aggregated over the run. The FLOP estimate excludes recomputation,
routing, and non-GEMM operations, but their execution costs remain in the measured
step time. This metric is not a hardware-counter measurement or model FLOPs
utilization (MFU).

\begin{table}[H]
    \centering
    \caption{End-to-end training performance of Qwen3-4B under Top-16, Top-32,
    and Top-64 reference budgets. We compare fixed TopK routing with
    \methodname{}'s dynamic threshold routing at matched average retained-block
    budgets, using the same optimized KV-centric forward and adapted backward
    implementations. Avg. CE and Last-10 CE
    denote the average training cross-entropy over all 100 steps and the final
    10 steps, respectively. Step (s) is the mean end-to-end training iteration
    time in seconds; TFLOPS/GPU reports effective sparse-computation throughput.
    Blocks denotes the average number of retained KV blocks per query. Speedup
    is the ratio of fixed-TopK to dynamic-threshold step times.}
    \label{tab:training-performance}
    \small
    \setlength{\tabcolsep}{4pt}
    \renewcommand{\arraystretch}{1.12}
    \begin{tabular}{llrrrrrr}
        \toprule
        \rowcolor{lmHeader}
        Budget & Method & Avg. CE $\downarrow$ & Last-10 CE $\downarrow$
            & Step (s) $\downarrow$ & TFLOPS/GPU $\uparrow$ & Blocks & Speedup \\
        \midrule
        16 & TopK & 2.2147 & 2.2756 & 22.280 & 162.48 & 15.883 & 1.000$\times$ \\
        16 & \methodname{} & 2.2222 & 2.2780 & 22.361 & 161.96 & 15.933 & 0.996$\times$ \\
        \midrule
        32 & TopK & 2.1909 & 2.2595 & 39.558 & 103.24 & 31.516 & 1.000$\times$ \\
        32 & \methodname{} & 2.1967 & 2.2615 & 32.284 & 126.41 & 31.405 & 1.225$\times$ \\
        \midrule
        64 & TopK & 2.1802 & 2.2531 & 85.645 & 58.26 & 62.031 & 1.000$\times$ \\
        64 & \methodname{} & 2.1810 & 2.2507 & 53.470 & 94.58 & 62.295 & 1.602$\times$ \\
        \bottomrule
    \end{tabular}
\end{table}

\paragraph{Training results.}
Table~\ref{tab:training-performance} summarizes end-to-end training performance
across three retention budgets. The reported average block counts differ by less
than 0.5\% between fixed TopK and dynamic threshold routing. At Top-16, the two
configurations achieve nearly identical training speed (0.996$\times$). With the
same execution backend, dynamic routing achieves 1.225$\times$ and 1.602$\times$
higher end-to-end training throughput at Top-32 and Top-64, respectively. Both
the average and final-10-step training CE remain close to the corresponding
fixed-TopK results. These findings demonstrate the applicability of \methodname{}
to training and the increasing benefit of dynamic routing at larger retention
budgets over the evaluated run.
\FloatBarrier

\FloatBarrier
\Needspace{110pt}
\subsection{System Efficiency and Ablation Studies}
\label{sec:exp:system-efficiency}

\FloatBarrier
\subsubsection{Cumulative Implementation Ablation}
\label{sec:exp:ablation:cumulative-body}

Adding one component at a time to FlashMoBA Top-$k$ gives the cumulative
device-path latency of Table~\ref{tab:cumulative-implementation-ablation}, each
variant timed on the same captured BF16 Q/K/V with routing, CSC construction,
attention and merge included. The budget exposes the scaling difference: the
final path is 1.56 to 1.65$\times$ faster than FlashMoBA at Top-$16$ and 2.16 to
2.56$\times$ at Top-$32$, because ordered Top-$k$ maintenance grows with $k$
while the threshold stage applies one scalar test per recomputed score.

\begingroup
\setlength{\parskip}{5pt}
\setlength{\intextsep}{6pt}
\captionsetup{skip=5pt}
\begin{table}[H]
    \centering
\caption{Cumulative device-path latency (ms): FlashMoBA, specialized
    unsplit KV kernel, split-$4096$ worklist, SM90 pipeline, and matched-density
    threshold routing. The first four variants share saved Top-$k$ CSC metadata.
    Speedup compares the final variant with FlashMoBA.}

    \label{tab:cumulative-implementation-ablation}
    \small
    \setlength{\tabcolsep}{2.2pt}
        \begin{tabular}{llrrrrrrr}
        \toprule
        \rowcolor{lmHeader}
        Model & Length & $k$ & FlashMoBA & +KV & +Split & +Pipeline & +Threshold & Speedup \\
        \midrule
        Qwen3-4B & 32K  & 16 & 25.529 & 22.091 & 20.564 & 16.520 & 15.632 & 1.63$\times$ \\
        Qwen3-4B & 128K & 16 & 137.243 & 104.740 & 97.489 & 81.813 & 83.284 & 1.65$\times$ \\
        \innergqa{} & 32K  & 16 & 24.543 & 21.574 & 19.961 & 15.866 & 15.786 & 1.56$\times$ \\
        \innergqa{} & 128K & 16 & 131.002 & 102.606 & 95.244 & 79.751 & 82.249 & 1.59$\times$ \\
        \midrule
        Qwen3-4B & 32K  & 32 & 56.083 & 48.102 & 47.797 & 37.821 & 24.291 & 2.31$\times$ \\
        Qwen3-4B & 128K & 32 & 312.387 & 262.775 & 261.526 & 221.020 & 121.845 & 2.56$\times$ \\
        \innergqa{} & 32K  & 32 & 52.660 & 45.178 & 45.107 & 35.154 & 24.380 & 2.16$\times$ \\
        \innergqa{} & 128K & 32 & 287.236 & 243.851 & 242.549 & 202.464 & 120.512 & 2.38$\times$ \\
        \bottomrule
    \end{tabular}
\end{table}

\paragraph{Timing scopes and stage breakdown.}
The fixed Top-$k$ variants share saved CSC metadata; the threshold variant uses
a separately saved pattern with matched valid pair counts. Adjacent variants
also change kernel families; single-factor tests follow below. The replay and
validation protocol appears in Appendix~\ref{sec:exp:ablation:cumulative}.

Fig.~\ref{fig:cumulative-attention-path} complements the cumulative table with a
stage breakdown against FlashMoBA and dense FA3 at 32K and 128K. Complete
attention device-path totals include pooling, routing/CSC construction,
reverse-index work, attention, and merge; isolated kernel timings exclude the
surrounding stages. Full-model generation loops in Sec.~\ref{sec:exp:main:speed}
also include other model computation and dense decoding. Replay speedups are
therefore not full-model inference speedups.
At 32K, LightningMoBA's complete attention path remains slower than dense FA3
(15.8 vs. 13.0 ms) due to sparse routing and metadata overheads, whereas at
128K, the reduction in attention computation outweighs these costs, yielding
a 2.56$\times$ speedup over FA3 (83.2 vs. 213.2 ms).

\begin{figure}[H]
    \centering
    \includegraphics[width=0.95\linewidth]{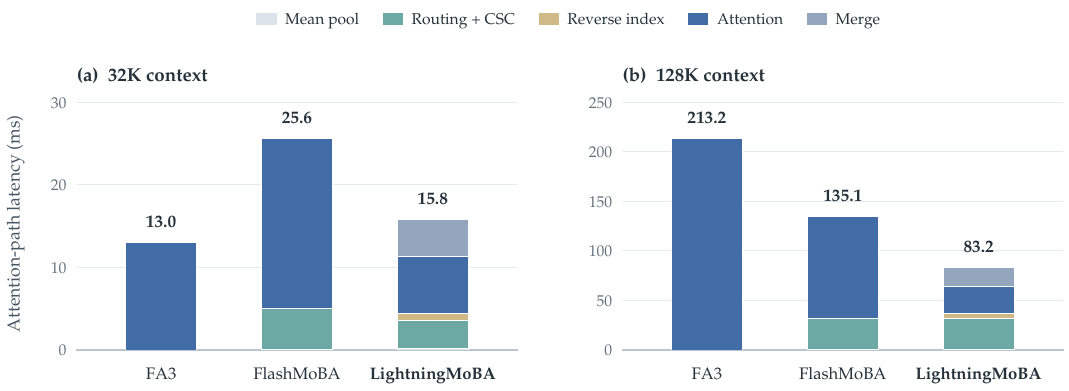}
    \caption{Complete forward attention-path timing breakdown for Qwen3-4B at
    (a) 32K and (b) 128K on one SM90 GPU. Each panel compares dense FA3,
    FlashMoBA Top-$16$, and the final LightningMoBA configuration. The panels
    use separate latency axes; sparse bars include pooling, routing/CSC
    construction, attention, reverse-index work, and merge.}
    \label{fig:cumulative-attention-path}
\end{figure}
\endgroup
\FloatBarrier

\FloatBarrier
\subsubsection{Routing Efficiency and End-to-End Impact}
\label{sec:exp:ablation:routing-body}
\label{sec:same-backend-routing}

Isolating routing cost shows why the budget matters. Against FlashMoBA's
ordered Top-$k$ candidate set at matched realized pair counts, threshold routing
is faster by 3.15 to 4.13$\times$ at $k=32$ and 6.99 to 11.25$\times$ at $k=64$,
and the gap widens with $k$ (Table~\ref{tab:routing-full-k-sweep}).

\begin{table}[htbp]
    \centering
    \caption{Complete routing sweep including CSC construction. Each group reports
    Top-$k$ and threshold latency (ms) with their ratio; each threshold setting
    matches the valid CSC pair count within 0.05\%. Timings use CUDA Graph replay.}

    \label{tab:routing-full-k-sweep}
    \small
    \setlength{\tabcolsep}{2.5pt}
        \resizebox{0.8\linewidth}{!}{\begin{tabular}{llrrr rrr rrr}
        \toprule
        \rowcolor{lmHeader}
        & & \multicolumn{3}{c}{$k=16$} & \multicolumn{3}{c}{$k=32$} & \multicolumn{3}{c}{$k=64$} \\
        \rowcolor{lmHeader}
        Model & Length & Top-$k$ & Thresh. & Speedup & Top-$k$ & Thresh. & Speedup & Top-$k$ & Thresh. & Speedup \\
        \midrule
        Qwen3-4B & 16K  & 1.86 & 1.40 & 1.33$\times$ & 6.32 & 1.91 & 3.31$\times$ & 20.63 & 2.88 & 7.17$\times$ \\
        Qwen3-4B & 32K  & 4.06 & 3.39 & 1.19$\times$ & 17.93 & 4.34 & 4.13$\times$ & 64.01 & 6.02 & 10.64$\times$ \\
        Qwen3-4B & 64K  & 10.25 & 10.36 & 0.99$\times$ & 47.47 & 12.65 & 3.75$\times$ & 177.28 & 17.26 & 10.27$\times$ \\
        Qwen3-4B & 128K & 30.70 & 32.28 & 0.95$\times$ & 135.33 & 36.81 & 3.68$\times$ & 514.42 & 45.74 & 11.25$\times$ \\
        \midrule
        \innergqa{} & 16K  & 1.66 & 1.35 & 1.23$\times$ & 5.58 & 1.77 & 3.15$\times$ & 19.22 & 2.75 & 6.99$\times$ \\
        \innergqa{} & 32K  & 3.52 & 3.26 & 1.08$\times$ & 15.28 & 4.12 & 3.71$\times$ & 54.65 & 6.22 & 8.79$\times$ \\
        \innergqa{} & 64K  & 10.01 & 9.68 & 1.03$\times$ & 44.82 & 11.62 & 3.86$\times$ & 159.34 & 15.99 & 9.96$\times$ \\
        \innergqa{} & 128K & 28.57 & 31.85 & 0.90$\times$ & 116.96 & 35.48 & 3.30$\times$ & 416.01 & 44.56 & 9.34$\times$ \\
        \bottomrule
    \end{tabular}}
\end{table}

These CUDA Graph timings include scoring, selection, and CSC construction, but
exclude mean pooling and attention. They therefore characterize the router, not
the complete device path or full-model inference. Each model--length--budget
configuration is calibrated outside timing to match attention-visible pair counts;
Appendix~\ref{sec:exp:ablation:routing-cost} details this protocol and the separate
decision-stage measurement. At Top-$16$, the complete-router advantage varies
with context length and reverses at 128K in both models; larger budgets favor
threshold selection in every measured case.

\paragraph{Controlled end-to-end impact.}

Three configurations distinguish execution from routing: FlashMoBA uses fixed
Top-$k$ with its original execution backend; optimized fixed Top-$k$ uses the
\methodname{} backend; and \methodname{} uses dynamic threshold routing with
that same optimized backend. Table~\ref{tab:dynamic-vs-topk} and
Fig.~\ref{fig:quality-latency-routing} in Appendix~\ref{sec:add-figures}
compare the latter two at matched average
retention budgets of 16, 32, and 64 on Qwen3-4B and \innergqa{} at 32K, 64K, and
128K. Every entry aggregates the full 6,500-example RULER suite at tensor
parallelism eight and batch size one. Both use the same KV-centric execution,
fanout-aware splitting, and runtime pipelining; times cover the complete
generation loop, not isolated routing or attention kernels. These fixed Top-$k$
results are therefore distinct from the original FlashMoBA baseline in
Table~\ref{tab:main-e2e}.

For corresponding measurements on the same model, length, and budget, let
$t_{\mathrm{F}}$, $t_{\mathrm{K}}$, and $t_{\mathrm{D}}$ denote the complete
runtimes of original FlashMoBA, optimized fixed Top-$k$, and dynamic routing.
The total speedup factors as
$t_{\mathrm{F}}/t_{\mathrm{D}}=(t_{\mathrm{F}}/t_{\mathrm{K}})
(t_{\mathrm{K}}/t_{\mathrm{D}})$: an execution-backend ratio times a routing-policy
ratio. These are ratios of measured complete runtimes, not independently additive
kernel-time contributions; the identity does not apply across different timing
protocols or unmatched configurations.

\begin{table}[H]
    \centering
    \caption{Controlled end-to-end comparison of fixed Top-$k$ and dynamic
    threshold routing under the identical optimized \methodname{} execution
    backend. Results cover Qwen3-4B and \innergqa{} at 32K, 64K, and 128K, with
    matched average retention budgets of 16, 32, and 64. Both strategies use the
    same KV-centric execution, fanout-aware splitting, and runtime pipelining.
    RULER scores are reported in percentages, generation times in minutes, and
    speedups are calculated as Fixed / Dynamic.}
    \label{tab:dynamic-vs-topk}
    \small
    \setlength{\tabcolsep}{7pt}
    \begin{tabular}{llrrrrrr}
        \toprule
        \rowcolor{lmHeader}
        Model & Length & Budget & \multicolumn{2}{c}{RULER score}
            & \multicolumn{2}{c}{E2E (min)} & Speedup \\
        \rowcolor{lmHeader}
        & & & Fixed & Dynamic & Fixed & Dynamic & \\
        \midrule
        Qwen3-4B & 32K & 16 & 70.48 & 79.93 & 123.52 & 114.33 & 1.08$\times$ \\
        & 32K & 32 & 78.30 & 87.87 & 150.54 & 131.43 & 1.15$\times$ \\
        & 32K & 64 & 86.06 & 91.13 & 248.04 & 213.02 & 1.16$\times$ \\
        \addlinespace[2pt]
        & 64K & 16 & 59.84 & 53.41 & 143.34 & 153.10 & 0.94$\times$ \\
        & 64K & 32 & 68.82 & 65.36 & 219.69 & 192.78 & 1.14$\times$ \\
        & 64K & 64 & 76.46 & 81.19 & 369.09 & 299.83 & 1.23$\times$ \\
        \addlinespace[2pt]
        & 128K & 16 & 45.57 & 46.51 & 247.79 & 251.57 & 0.98$\times$ \\
        & 128K & 32 & 54.31 & 60.76 & 392.99 & 353.73 & 1.11$\times$ \\
        & 128K & 64 & 63.30 & 69.25 & 687.21 & 461.59 & 1.49$\times$ \\
        \midrule
        \innergqa{} & 32K & 16 & 74.16 & 70.82 & 95.56 & 103.92 & 0.92$\times$ \\
        & 32K & 32 & 82.68 & 85.17 & 121.27 & 109.92 & 1.10$\times$ \\
        & 32K & 64 & 88.69 & 90.46 & 215.80 & 183.31 & 1.18$\times$ \\
        \addlinespace[2pt]
        & 64K & 16 & 58.20 & 55.72 & 142.82 & 122.45 & 1.17$\times$ \\
        & 64K & 32 & 65.39 & 65.19 & 239.76 & 189.85 & 1.26$\times$ \\
        & 64K & 64 & 72.34 & 73.19 & 379.95 & 280.54 & 1.35$\times$ \\
        \addlinespace[2pt]
        & 128K & 16 & 34.84 & 35.40 & 234.54 & 253.10 & 0.93$\times$ \\
        & 128K & 32 & 39.91 & 39.31 & 408.33 & 369.00 & 1.11$\times$ \\
        & 128K & 64 & 46.99 & 49.47 & 803.53 & 515.06 & 1.56$\times$ \\
        \bottomrule
    \end{tabular}
\end{table}

\paragraph{Budget-dependent routing benefits.}
Table~\ref{tab:dynamic-vs-topk} and Fig.~\ref{fig:quality-latency-routing}
demonstrate that the independent contribution of dynamic routing depends on the
retention budget. At Top-16, dynamic routing is slower in four of the six
configurations and does not consistently improve RULER accuracy, indicating that
the primary end-to-end acceleration reported at this budget arises from the
optimized execution backend.

At Top-32, dynamic routing reduces end-to-end generation time across all six
configurations, achieving 1.10--1.26$\times$ speedups, while accuracy varies across
models and context lengths. At Top-64, dynamic routing achieves simultaneous
latency and quality improvements in all six configurations, delivering
1.16--1.56$\times$ speedups and RULER accuracy gains of 0.85--5.95 percentage points.

These results complement the routing and CSC construction measurements in
Table~\ref{tab:routing-full-k-sweep}, where the advantage of threshold-based
selection increases substantially with the retention budget. Together, the
controlled comparisons distinguish the contribution of the execution backend from
the additional benefits of dynamic routing at larger budgets in the measured
models and context lengths.

\FloatBarrier

\FloatBarrier
\Needspace{205pt}
\subsubsection{KV-Centric Ownership}
\label{sec:exp:ablation:ownership-body}

On identical Top-$16$ metadata and captured tensors, column ownership cuts
attention-stage latency by 18.9\% at 32K and 33.8\% at 128K (16.9\% and 31.0\% on
\innergqa{}) by removing repeated K/V loads (Table~\ref{tab:kv-ownership-ablation}).
Validation replays of the same routing give maximum absolute and relative $L_2$
output errors below $1.6\times10^{-2}$ and $9.4\times10^{-4}$; the validation
protocol is described in Appendix~\ref{sec:exp:ablation:kv-ownership}.
KV-centric units serve whole columns, so each unit's task size follows its
column's fanout, and fanout is strongly skewed: the largest column
carries 64.5$\times$ the mean and the median only 0.064$\times$
(Fig.~\ref{fig:fanout-halfcol}). Sec.~\ref{sec:exp:ablation:fanout} bounds the
resulting tail by splitting.

\begin{table}[H]
    \centering
    \caption{Ownership ablation on identical Top-$16$ routing metadata and
    complete 32/8-head model tensors; latencies are attention-stage
    milliseconds. $\Lambda(M)$ is the mean number of query-centric units touching
    a selected $(h,j)$ column; $\mu(M)$ and $\mu_{\mathrm{KV}}$ are the mean
    selected rows served per query-centric and KV-centric unit.}
    \label{tab:kv-ownership-ablation}
    \small
    \setlength{\tabcolsep}{4pt}
        \resizebox{0.75\linewidth}{!}{\begin{tabular}{llrrrrrr}
        \toprule
        \rowcolor{lmHeader}
        Model & Length & Query & KV unsplit & Speedup & $\Lambda(M)$ & $\mu(M)$ & $\mu_{\mathrm{KV}}$ \\
        \midrule
        Qwen3-4B & 32K  & 20.056 & 16.263 & 1.23$\times$ & 13.66 & 145.50 & 1988.00 \\
        Qwen3-4B & 128K & 100.843 & 66.727 & 1.51$\times$ & 20.91 & 97.24 & 2033.00 \\
        \innergqa{} & 32K  & 19.644 & 16.316 & 1.20$\times$ & 10.88 & 182.72 & 1988.00 \\
        \innergqa{} & 128K & 96.207 & 66.380 & 1.45$\times$ & 16.03 & 126.79 & 2032.99 \\
        \bottomrule
    \end{tabular}}
\end{table}

\FloatBarrier
\subsubsection{Fanout-Aware Splitting}
\label{sec:exp:ablation:fanout}

The decomposition in Sec.~\ref{sec:method:fanout} caps the longest work item to
improve scheduling parallelism, at the cost of additional work items and K/V
loads. We isolate this trade-off using the captures of
Appendix~\ref{sec:exp:ablation:kv-ownership}: one Top-$16$ CSC pattern is fixed per
capture, and only the selected-row cap $T\in\{\infty,8192,4096,2048\}$ varies.
All variants use the same flat-worklist kernel and merge, without work-item
sorting or query-gather pipelining. The $T=\infty$ control assigns one item per
nonempty column through this same path, unlike the specialized unsplit kernel
in the ownership experiment. Timing includes attention and merge; routing,
reverse-index construction, and host-built worklists are outside timing, so the
measurements exclude production device-side worklist construction costs.

\begin{table}[htbp]
    \centering
    \caption{Fanout-aware splitting. Latencies include the
    attention kernel and merge, in ms, on the same CSC pattern within each
    row. Columns $\rightarrow$ items gives nonempty $(h,j)$ columns and work
    items at $T=4096$; their ratio is $\Lambda_{4096}$. Max tiles gives
    $\max w$ for $T=\infty\rightarrow4096$ with $m_{\mathrm{MMA}}=64$.
    Speedup compares unsplit with $T=4096$.}
    \label{tab:fanout-split-ablation}
    \small
    \setlength{\tabcolsep}{3pt}
        \resizebox{\linewidth}{!}{\begin{tabular}{llrrrrrrr}
        \toprule
        \rowcolor{lmHeader}
        Model & Length & Columns $\rightarrow$ items & Max tiles & Unsplit & $T=8192$ & $T=4096$ & $T=2048$ & Speedup \\
        \midrule
        Qwen3-4B & 32K  & $8192\rightarrow8960$   & $512\rightarrow64$ & 17.406 & 15.176 & 14.795 & 14.683 & 1.18$\times$ \\
        Qwen3-4B & 128K & $32768\rightarrow35040$ & $2048\rightarrow64$ & 71.189 & 59.832 & 59.901 & 59.755 & 1.19$\times$ \\
        \innergqa{} & 32K  & $8192\rightarrow8467$   & $512\rightarrow64$ & 17.464 & 15.176 & 14.803 & 14.664 & 1.18$\times$ \\
        \innergqa{} & 128K & $32736\rightarrow34049$ & $2046\rightarrow64$ & 71.053 & 59.680 & 59.666 & 59.539 & 1.19$\times$ \\
        \bottomrule
    \end{tabular}}
\end{table}

Splitting the actual all-pair columns at $T=4096$ reduces attention-stage
latency by 15.0 to 16.0\% (Table~\ref{tab:fanout-split-ablation}).
The largest work item falls from 512 tiles at 32K and 2,046 to 2,048 tiles
at 128K to 64 tiles, while structural K/V load multiplicity
$\Lambda_{4096}$ rises to 1.034 to 1.094. This count is not measured physical
memory traffic. Smaller caps further limit the longest item but create more
items and repeated K/V loads; the default $T=4096$ balances these costs.
The broader structural trade-off is illustrated in Fig.~\ref{fig:fanout}, and
optional-only fanout diagnostics are retained in
Appendix~\ref{sec:exp:ablation:fanout-diagnostics}
(Fig.~\ref{fig:fanout-split-tail}). Similar speedups despite different optional
tails indicate that optional fanout alone does not predict the gain: both
all-pair patterns also contain full-length forced sink columns.
\FloatBarrier

\FloatBarrier
\subsubsection{SM90 Query-Gather Pipelining}
\label{sec:exp:ablation:pipeline}

With Top-$16$ routing, $T=4096$ partitioning, and the same TMA/WGMMA
configuration and shared-memory allocation, the two-stage query-gather pipeline
yields 2.53 to 2.65$\times$ speedup and zero measured output error on 8,192
sampled CSC rows per case (Table~\ref{tab:sm90-pipeline-ablation}).

\begin{table}[htbp]
    \centering
    \begin{minipage}[b]{0.46\linewidth}
        \centering
        \small
        \setlength{\tabcolsep}{2.5pt}
        \begin{tabular}{llrrr}
            \toprule
            \rowcolor{lmHeader}
            Model & Len. & Serial & Piped & Speedup \\
            \midrule
            Qwen3-4B & 32K & 17.642 & 6.868 & 2.57$\times$ \\
            Qwen3-4B & 128K & 71.412 & 26.951 & 2.65$\times$ \\
            \innergqa{} & 32K & 17.053 & 6.741 & 2.53$\times$ \\
            \innergqa{} & 128K & 69.586 & 26.806 & 2.60$\times$ \\
            \bottomrule
        \end{tabular}
        \caption{SM90 query-gather pipeline ablation with the same Top-$16$
        routing, $T=4096$ work partition and CUDA Graph timing scope; latencies
        are milliseconds. ``Serial'' denotes serial Q gathering in the same
        kernel.}
        \label{tab:sm90-pipeline-ablation}
    \end{minipage}\hfill
    \begin{minipage}[b]{0.50\linewidth}
        \centering
        \includegraphics[width=\linewidth]{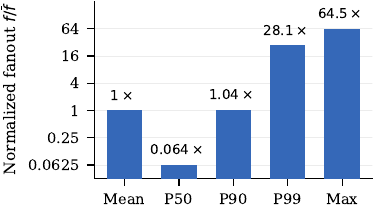}
        \captionof{figure}{Column fanout imbalance across 4,096 nonempty
        query-head/KV-block columns of a high-tail \innergqa{} TP8 RULER-128K
        prefill trace, normalized by the mean.}
        \label{fig:fanout-halfcol}
    \end{minipage}
\end{table}

\FloatBarrier
\subsection{Long-Context Quality Analysis}
\label{sec:exp:quality-granularity}
\suppressfloats[t]

\paragraph{Task-level quality diagnosis.}
We examine long-context quality using all 13 RULER subtasks across five models at
64K and 128K under the Top-16 reference budget. The complete results against the
original FlashMoBA baseline are reported in
Appendix~\ref{sec:task-level-ruler}
(Tables~\ref{tab:ruler-tasks-64k} and~\ref{tab:ruler-tasks-128k}). At 64K,
variable tracking (VT) accuracy decreases by an average of 25.77 percentage points
across the five models, accounting for approximately 55.3\% of the reduction in
the 13-task macro-average RULER score. At 128K, the average VT difference narrows
to $-8.10$ percentage points. VT involves multi-hop dependencies and may therefore
be particularly sensitive to the selection of intermediate evidence blocks.

\paragraph{Retention-budget sensitivity.}
We first investigate whether retaining more information mitigates the observed
quality degradation. Table~\ref{tab:ruler-budget-categories} compares
\methodname{} with the original FlashMoBA baseline under Top-16, Top-32, and
Top-64 reference budgets on Qwen3-4B and \innergqa{} at 64K and 128K. This is a
full-system comparison against FlashMoBA, distinct from the same-backend
fixed-versus-dynamic routing comparison in Sec.~\ref{sec:same-backend-routing}.

\begin{table}[!htbp]
    \centering
    \caption{Long-context RULER retention-budget sensitivity on Qwen3-4B and
    \innergqa{} at 64K and 128K under Top-16, Top-32, and Top-64 reference budgets.
    Entries are \methodname{} $-$ FlashMoBA accuracy differences in percentage
    points. Retrieval, Multi-hop Tracing (VT), Aggregation, and QA comprise 8, 1, 2,
    and 2 subtasks; Overall is the 13-task macro-average.}
    \label{tab:ruler-budget-categories}
    \small
    \setlength{\tabcolsep}{4pt}
    \renewcommand{\arraystretch}{1.10}
    \begin{tabular}{lllrrrrr}
        \toprule
        \rowcolor{lmHeader}
        Model & Length & Budget & Retrieval & Multi-hop Tracing (VT)
            & Aggregation & QA & Overall \\
        \midrule
        Qwen3-4B & 64K & 16 & $-2.51$ & $-36.96$ & $-4.26$ & $-4.07$ & $-5.67$ \\
        & & 32 & $-2.25$ & $-23.00$ & $+5.92$ & $-6.69$ & $-3.27$ \\
        & & 64 & $+7.26$ & $-0.40$ & $-0.87$ & $+2.75$ & $+4.73$ \\
        \addlinespace[2pt]
        & 128K & 16 & $+3.04$ & $-15.16$ & $-0.79$ & $-2.78$ & $+0.15$ \\
        & & 32 & $+11.18$ & $-18.56$ & $-0.82$ & $+4.79$ & $+6.06$ \\
        & & 64 & $+8.79$ & $+1.44$ & $-1.86$ & $+4.63$ & $+5.95$ \\
        \midrule
        \innergqa{} & 64K & 16 & $-0.30$ & $-23.76$ & $+3.00$ & $+0.89$ & $-1.41$ \\
        & & 32 & $+5.51$ & $-34.76$ & $-0.93$ & $-1.93$ & $+0.27$ \\
        & & 64 & $+5.19$ & $-22.84$ & $-1.81$ & $-1.97$ & $+0.85$ \\
        \addlinespace[2pt]
        & 128K & 16 & $+2.67$ & $-12.00$ & $-2.20$ & $+0.62$ & $+0.48$ \\
        & & 32 & $-0.62$ & $-5.80$ & $+0.83$ & $+1.03$ & $-0.54$ \\
        & & 64 & $+4.12$ & $-2.72$ & $-1.39$ & $+2.40$ & $+2.48$ \\
        \bottomrule
    \end{tabular}
\end{table}

Increasing the budget from Top-16 to Top-64 improves the Overall difference in
all four configurations, with \methodname{} outperforming FlashMoBA overall at
Top-64. At 64K, the difference improves from $-5.67$ to $+4.73$ percentage points
for Qwen3-4B and from $-1.41$ to $+0.85$ for \innergqa{}; the Qwen3-4B VT
difference narrows from $-36.96$ to $-0.40$ points. Gains vary across task categories.

\FloatBarrier
\paragraph{Block-granularity sensitivity.}
As a complementary approach, we adapt \methodname{} to support block size 8 and
compare it with the default block size 128 on Qwen3-4B and \innergqa{} at 64K and
128K. Both configurations have matched average retained-token budgets, allowing
us to examine finer-grained routing without increasing the average number of
retained attention tokens. Table~\ref{tab:block-granularity} reports the resulting
quality--efficiency trade-off within \methodname{}.

\begin{table}[H]
    \centering
    \caption{Block-granularity sensitivity within \methodname{} on Qwen3-4B and
    \innergqa{} at 64K and 128K under matched average retained-token budgets.
    Accuracy differences are Block-8 $-$ Block-128 in percentage points;
    positive values indicate higher accuracy with Block-8. Latency ratio is
    $\mathrm{E2E}(\text{Block-8})/\mathrm{E2E}(\text{Block-128})$, using end-to-end
    generation time; values above one indicate that Block-8 is slower.}
    \label{tab:block-granularity}
    \small
    \setlength{\tabcolsep}{5pt}
    \renewcommand{\arraystretch}{1.34}
    \begin{tabular}{llrrrrrr}
        \toprule
        \rowcolor{lmHeader}
        Model & Length & $\Delta$VT & $\Delta$Retrieval & $\Delta$Aggregation
            & $\Delta$QA & $\Delta$Overall & Latency ratio \\
        \midrule
        Qwen3-4B & 64K & $+16.80$ & $+2.45$ & $+2.47$ & $-0.95$ & $+3.03$ & $2.32\times$ \\
        & 128K & $+10.20$ & $+3.47$ & $+0.54$ & $+5.90$ & $+3.91$ & $2.48\times$ \\
        \midrule
        \innergqa{} & 64K & $+17.00$ & $+3.90$ & $+3.80$ & $+11.36$ & $+6.04$ & $2.26\times$ \\
        & 128K & $+23.80$ & $+4.86$ & $+2.00$ & $+16.91$ & $+7.73$ & $2.25\times$ \\
        \bottomrule
    \end{tabular}
\end{table}

Block-8 improves VT accuracy by 10.20--23.80 percentage points and overall RULER
accuracy by 3.03--7.73 points across the four configurations, but incurs
2.25--2.48$\times$ the end-to-end latency of Block-128. Under mean-pooled block
scoring, unrelated tokens in a large block may dilute locally relevant evidence;
finer blocks may better distinguish that evidence. Increasing the retention budget
and refining block granularity thus offer complementary ways to improve
long-context quality.
Block-128 remains the default configuration to balance quality and end-to-end
efficiency.
\FloatBarrier

\section{Conclusion}
\label{sec:conclusion}
\method{} keeps MoBA's scoring function and retained-token softmax while
decoupling retention from task formation: threshold routing adapts retention to
each score distribution, KV-centric ownership and fanout-aware splitting form
reusable bounded tasks, and Hopper pipelining overlaps K/V movement, query
gathering and Tensor Core work. At matched Top-$16$ retention these mechanisms
accelerate end-to-end generation by 1.20 to 1.60$\times$ across five checkpoints
while increasing mean RULER score from 74.42 to 75.28; in controlled replays,
complete device-path speedup grows to 2.16 to 2.56$\times$ at Top-$32$, while
routing and CSC construction reach 6.99 to 11.25$\times$ at Top-$64$.
Same-backend comparisons show that execution optimizations account for most
Top-$16$ acceleration, while dynamic routing provides increasing benefits at
larger budgets, reaching 1.16 to 1.56$\times$ end-to-end speedups at Top-$64$ on
the evaluated two models. End-to-end Qwen3-4B training with the adapted FlashMoBA
backward implementation achieves a 1.602$\times$ speedup at Top-$64$, with
closely aligned cross-entropy losses over the evaluated 100-step run. These
results support the applicability of the design to both inference and training,
while distinguishing execution-backend improvements from routing-policy gains.

\FloatBarrier

\section*{Ethics Statement}
The study is algorithmic and systems-level: it changes the selection rule, task
formation and data movement of block-sparse attention and reports measurements
on public benchmarks. No human subjects, human annotations, or personally
identifiable or sensitive data are involved, and the study does not target a
high-risk application domain. We foresee no immediate ethical concerns arising
from this line of systems work.

\section*{Reproducibility Statement}
The threshold rule, ownership model, task decomposition and pipeline schedule
are specified in Section~\ref{sec:method}, with the streaming metadata
construction and the forward-pass pseudocode deferred to
Appendix~\ref{sec:router-details} and Algorithm~\ref{alg:kvcentric}.
Section~\ref{sec:exp:setup} lists the checkpoints, context lengths and benchmark.
End-to-end inference and training are reported in
Sections~\ref{sec:exp:main:speed} and \ref{sec:exp:training-performance}, respectively;
external sparse baselines are reported in Appendix~\ref{sec:sparse-baselines}.
System ablations and the complete attention path breakdown appear in
Sec.~\ref{sec:exp:system-efficiency}. Long-context
quality is analyzed in Sec.~\ref{sec:exp:quality-granularity}. The appendix
retains detailed protocols, profiler counters, rank-imbalance analyses, and full
per-task results. Proofs of both
lemmas appear in Appendix~\ref{sec:proofs}. Four of the five checkpoints are
publicly released; \innergqa{} is an internal model that cannot be distributed.
The code repository is \url{https://github.com/Tencent/LightningMoBA}.

\setcitestyle{square,numbers,comma}

\appendix

\section{Additional Methodology}
\subsection{Complexity Analysis}
\label{sec:method:complexity}

A call has four stages, and under the reference design of
Sec.~\ref{sec:method:prelim} they scale as
\begin{equation}
    \mathcal{O}(N^2 d / B)\ \text{(scoring)},\quad
    \mathcal{O}(N^2 k / B)\ \text{(selection)},\quad
    \mathcal{O}(Z)\ \text{(index)},\quad
    \mathcal{O}(Z B d)\ \text{(attention)},
    \label{eq:cost}
\end{equation}
where scoring and selection remain dense over the candidate blocks, while the
emitted metadata volume and the attention work scale with the realized pattern.
The candidate-set update within each query is comparison-bound and involves
sequential dependence, with cost increasing in both $n$ and $k$, since the
reference router maintains an ordered candidate set.
The three design axes act on different terms of Eq.~\eqref{eq:cost}; beyond
arithmetic work, the proposed execution also changes memory-traffic
multiplicity and scheduling granularity.
Table~\ref{tab:complexity} summarizes each component against its corresponding
reference design, which differs by component.

\begin{table}[h]
    \centering
    \caption{Arithmetic, movement, and scheduling costs for one
    attention call. Each row states its own
    reference design, and data-dependent quantities appear as realized rather
    than bounded.}
    \label{tab:complexity}
    \small
    \begin{tabular}{lll}
        \toprule
        \rowcolor{lmHeader}
        Component & Reference design & Proposed design \\
        \midrule
        Scoring
            & one pass, $\mathcal{O}(N^2 d / B)$
            & two passes, same asymptotic order \\
        Selection
            & ordered $k$-set, $\mathcal{O}(N^2 k / B)$
            & threshold test, $\mathcal{O}(N^2 / B)$ \\
        Index construction
            & count + prefix-sum + scatter, $\mathcal{O}(Z)$
            & scan + emit, $\mathcal{O}(N n + Z)$ \\
        Attention arithmetic
            & $\mathcal{O}(Z B d)$
            & unchanged, $\mathcal{O}(Z B d)$ \\
        \midrule
        K/V movement
            & $\mathcal{O}(\Lambda(M)\, C B d)$
            & $\mathcal{O}(\Lambda_T\, C B d)$ \\
        Longest work item
            & $\max_{h,j} w_{h,j}$
            & $\leq \lceil T/m_{\mathrm{MMA}}\rceil$ \\
        \bottomrule
    \end{tabular}
\end{table}

Threshold selection removes that $k$-dependent ordered-set maintenance, so the
improvement is against the procedure Eq.~\eqref{eq:cost} prices rather than
against Top-$k$ in general, while requiring two score passes to obtain the
per-query maximum before any comparison. Neither pass materializes a
floating-point score matrix. Index construction additionally scans the packed
candidate mask, giving $\mathcal{O}(Nn+Z)$ work at candidate granularity.
Because threshold routing does not impose a fixed per-query retention budget,
$Z$ is data-dependent and can reach $\Theta(N^2/B)$ in the dense-block limit,
so attention complexity must be compared at matched realized sparsity;
Sec.~\ref{sec:exp:ablation:routing-body} provides this controlled comparison.

KV-centric ownership and fanout splitting leave the $\mathcal{O}(ZBd)$ arithmetic
unchanged; they instead reduce fetch multiplicity and bound the longest serial
work item, trading reuse against parallelism. Every bound here is per attention call and does not
by itself determine end-to-end latency, which Sec.~\ref{sec:exp} measures.

\section{Additional Results and Discussion}
\subsection{Replay Protocols and Supplementary Diagnostics}
\label{sec:exp:ablation}
Unless otherwise stated, we replay captured BF16 tensors from Qwen3-4B and
\innergqa{} at 32K and 128K on a single rank, retaining the complete
32-query-head/8-KV-head shape with head dimension 128 and block size 128.
Comparisons use identical causal and compulsory-block rules. Latencies are
medians of 50 measurements after 10 warmups; each subsection specifies its
timing scope and CUDA Graph usage. These device replays exclude host-side
work and model inference, and remove the cross-rank synchronization present
in the TP8 main experiments.
Execution-only comparisons hold routing metadata fixed; routing-policy
comparisons match the realized number of selected query-block pairs.
The following protocols supplement the cumulative and single-component results
in Sec.~\ref{sec:exp:system-efficiency}; optional-only fanout diagnostics remain
here.

\subsubsection{Cumulative Implementation Ablation Protocol}
\label{sec:exp:ablation:cumulative}

Table~\ref{tab:cumulative-implementation-ablation} adds KV-centric ownership,
fanout splitting, query-gather pipelining, and threshold routing to FlashMoBA.
At each Top-$k$ budget, the first four variants consume one identical saved
CSC pattern. Their timed replays execute the Top-$k$ router but feed the saved
pattern to downstream kernels, controlling run-to-run changes in column counts.
The threshold variant uses a separate saved pattern with valid pair count
matched within 0.05\%; selected pair identities may differ.

Timing includes pooling, routing, required reverse-index construction,
device-side worklist construction, attention, and merge. We use eager
CUDA events because the baseline's segmented-sort path is not graph-capturable
in this replay. These fixed-pattern device-path costs are distinct from live
router-to-attention invocations and the CUDA Graph kernel timings below.

Adjacent columns are implementation steps rather than single-switch causal
tests: +Split replaces the specialized unsplit kernel with a worklist kernel,
and +Pipeline also changes the kernel family and merge path.
Secs.~\ref{sec:exp:ablation:fanout} and
\ref{sec:exp:ablation:pipeline} isolate these effects within fixed kernels.
Sampled-row output validation compares the same-pattern Top-$k$ variants against
FlashMoBA; it excludes the threshold arm, which selects different pairs.
Results are discussed in Sec.~\ref{sec:exp:ablation:cumulative-body}.

\subsubsection{Routing Cost Across Budgets Protocol}
\label{sec:exp:ablation:routing-cost}

We compare the cost of producing attention-ready CSC metadata.
FlashMoBA scores candidates once while maintaining an ordered $k$-element
set; threshold routing scores them twice, first for the per-query maximum
and then for threshold selection and packed-mask updates.
This tests the extra score pass against ordered-set maintenance.

We replay captured Q and mean-pooled K at 16K, 32K, 64K, and 128K for
$k\in\{16,32,64\}$. A separate $\alpha$ is calibrated outside timing for
each model, length, and $k$, matching total attention-visible valid CSC pairs
within 0.05\% across all 24 comparisons. Compulsory blocks count toward this
total and occupy slots within FlashMoBA's fixed budget. Threshold CSC columns
use device-resident counts to delimit their valid prefixes.
CUDA Graph timings include scoring, selection, and CSC construction, excluding
mean pooling and attention. Results appear in
Sec.~\ref{sec:exp:ablation:routing-body} and Table~\ref{tab:routing-full-k-sweep}.

A separate 128K decision-stage measurement excludes final CSC offset/scatter
construction but retains scoring, selection, and output/count updates.
Here threshold routing uses one fixed $\alpha$ calibrated at Top-$16$,
rather than matching each $k$ separately. This protocol does not isolate
set-update instructions from other effects of kernel specialization.

\subsubsection{Ownership Ablation Protocol}
\label{sec:exp:ablation:kv-ownership}

We compare query-centric and KV-centric execution using identical Q/K/V and
Top-$16$ CSC metadata, with one sink block and a two-block local window.
The captures contain post-normalization, post-RoPE Q/K and corresponding V.
Qwen3-4B is captured with one tensor-parallel rank; \innergqa{}'s eight rank
captures are concatenated along the head dimension for the full-head replay.

The KV-centric variant assigns one CTA per $(h,j)$ column, using an unsplit
$64\times64$ kernel and online-softmax merge, with fanout splitting, query
splitting, and worklist scheduling disabled. Timing covers the query-centric
attention kernel or the KV-centric kernel plus its required merge; routing
and reverse-index construction are excluded from both.

Selected pairs and the attention definition are preserved, while accumulation
order changes. Validation replays compare maximum absolute and relative $L_2$
output errors on the same routing patterns. Results are reported in
Sec.~\ref{sec:exp:ablation:ownership-body} and Table~\ref{tab:kv-ownership-ablation}.

\subsubsection{Optional Fanout Diagnostics}
\label{sec:exp:ablation:fanout-diagnostics}

\begin{figure}[htbp]
    \centering
    \includegraphics[width=0.95\linewidth]{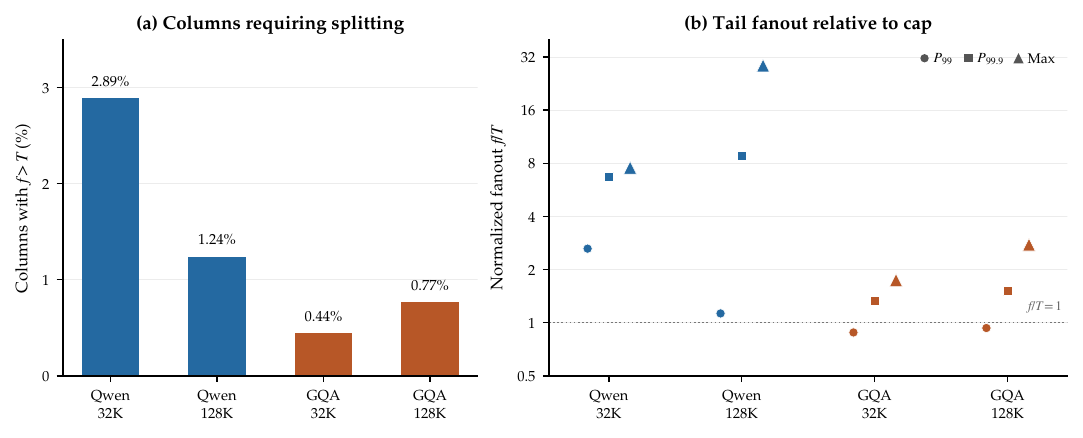}
    \caption{Selective splitting targets a small but potentially severe
    optional-route fanout tail. (a) Fraction of nonempty optional $(h,j)$
    columns with $f_{h,j}^{\mathrm{opt}}>T=4096$. (b) P99, P99.9, and maximum
    optional fanout normalized by $T$; the dotted line marks $f/T=1$.
    Qwen and \innergqa{} denote Qwen3-4B and \innergqa{}. Forced sink and local
    query-to-block pairs are excluded from both panels. Structural counts
    come from separate routing replays of the same captured Q/K inputs used
    for the latency comparison.}
    \label{fig:fanout-split-tail}
\end{figure}

The controlled splitting experiment and timing scope are reported in
Sec.~\ref{sec:exp:ablation:fanout} and Table~\ref{tab:fanout-split-ablation}.
Fig.~\ref{fig:fanout-split-tail} characterizes optional fanout
$f_{h,j}^{\mathrm{opt}}=|\{i:j\in\mathcal S_{h,i}\setminus\mathcal F_i\}|$,
excluding compulsory sink/local selections. Only 0.44\% to 2.89\% of nonempty
optional columns exceed $T=4096$, yet maximum optional fanout at 128K differs
by 10.3$\times$ between the captures (117,179 versus 11,337 rows).
This describes the sampled patterns, not a general model-architecture ranking.
All tested caps outperform unsplit; their latencies differ by at most 3.5\%
at 32K and 0.25\% at 128K relative to each case's fastest cap.
We use $T=4096$ thereafter. Validation replays give zero maximum absolute
and relative $L_2$ output errors.

\FloatBarrier
\subsection{Profiler and Tensor-Parallel Analysis}
\label{sec:exp:analysis}

The complete attention-path breakdown is reported in
Sec.~\ref{sec:exp:ablation:cumulative-body}
(Fig.~\ref{fig:cumulative-attention-path}). Here we retain supplementary profiler
counters and tensor-parallel rank-imbalance measurements.

\subsubsection{Pipeline Stall and Activity Analysis}
\label{sec:exp:analysis:pipeline}

We profile the attention kernel of the serial and pipelined variants from
Sec.~\ref{sec:exp:ablation:pipeline} on the two 128K captures using Nsight
Compute. These profiler runs do not supply the CUDA Graph timings.

With pipelining, long-scoreboard stall falls from about 63\% to below 8\%,
while tensor-pipe activity rises from about 7\% to 17\% and SM issue activity
from about 15\% to 36\% (Table~\ref{tab:sm90-pipeline-ncu}).
These whole-kernel counters support overlap of irregular Q gathers with
matrix work, but do not resolve producer/consumer timelines or attribute
the latency gain to individual warps. Counter definitions appear in the
caption.

\begin{table}[htbp]
    \centering
    \caption{Kernel-level Nsight Compute counters at 128K for the serial
    and pipelined variants of Sec.~\ref{sec:exp:ablation:pipeline}.
    Tensor-pipe and SM issue activity are percentages of peak sustained
    rate during unit-active cycles; long scoreboard is the percentage of
    active warps per cycle waiting on an L1TEX scoreboard dependency
    (local, global, surface, or texture access).}
    \label{tab:sm90-pipeline-ncu}
    \small
    \setlength{\tabcolsep}{4pt}
        \begin{tabular}{llrrr}
        \toprule
        \rowcolor{lmHeader}
        Model & Variant & Long scoreboard (\%) & Tensor pipe (\%) & SM issue (\%) \\
        \midrule
        Qwen3-4B & No pipeline & 63.33 & 7.33 & 15.05 \\
        Qwen3-4B & Pipelined & 7.87 & 17.08 & 36.01 \\
        \innergqa{} & No pipeline & 62.56 & 7.49 & 15.33 \\
        \innergqa{} & Pipelined & 7.67 & 17.14 & 36.03 \\
        \bottomrule
    \end{tabular}
\end{table}

\FloatBarrier
\Needspace{9\baselineskip}
\subsection{Discussion}
\label{sec:discussion}

\subsubsection{Architectural Discussion}
The design separates architecture-agnostic work organization from
Hopper-specific execution mechanisms. KV-centric ownership, fanout-aware
splitting, and reverse-index metadata determine how routed work is represented
and scheduled, independent of TMA or WGMMA. Our Hopper implementation maps the
regular K/V path to TMA, the irregular query-gather path to a dedicated
producer warp, and matrix computation to WGMMA, allowing data movement and
Tensor Core computation to overlap. Porting to another accelerator therefore
requires hardware support, or an alternative software mechanism, for
concurrent contiguous transfers, indexed gathers, and matrix operations, but
does not require changing the routing rule or ownership abstraction.
Consequently, the pipelining gains reported here are tied to Hopper, whereas
the routing and ownership analyses are hardware-independent.

\subsubsection{Impact on Tensor-Parallel Load Balance}
Fanout-aware splitting bounds the work assigned to one KV column within a
rank, but it does not redistribute work between tensor-parallel ranks. We
analyze an instrumented Qwen3-4B RULER-128K prefill profile with eight ranks,
each owning four query heads and one KV head. Excluding the first request,
12 requests and 36 layers provide 432 complete request-layer groups. For each
group, we divide the maximum rank-local value by the mean across ranks;
Table~\ref{tab:tp-rank-balance} aggregates these ratios. The observed rank-level
timing is consistent with workload imbalance: the rank with the most selected
query-block pairs is also the slowest K1 rank in 428 of 432 request-layer groups.

\begin{table}[htbp]
    \centering
    \caption{Observed TP rank imbalance on Qwen3-4B RULER-128K prefill.
    Ratios compare the maximum and mean of eight ranks within each of 432
    request-layer groups. K1 denotes KV-centric attention execution; the
    complete stage includes routing, scheduling, K1, and merge.}
    \label{tab:tp-rank-balance}
    \small
    \setlength{\tabcolsep}{7pt}
    \begin{tabular}{lrr}
        \toprule
        \rowcolor{lmHeader}
        Rank-local quantity & Mean max/mean & P90 max/mean \\
        \midrule
        Selected query-block pairs & 1.88$\times$ & 2.65$\times$ \\
        KV work items & 1.87$\times$ & 2.62$\times$ \\
        K1 latency & 1.86$\times$ & 2.61$\times$ \\
        Complete attention-stage latency & 1.39$\times$ & 1.71$\times$ \\
        \bottomrule
    \end{tabular}
\end{table}

The remaining K1 imbalance tracks rank-local workload even with within-rank
splitting enabled. The profile times individual prefill
stages with CUDA events and disables decode CUDA Graph for instrumentation;
these ratios diagnose a limitation of the current head placement, not an
end-to-end speedup estimate. Rank-aware budget control or cross-rank work
redistribution could reduce the slow-rank path, but the former can alter the
selected pattern and the latter incurs communication. We leave both as future
work rather than assuming a free TP balancing gain.

\Needspace{12\baselineskip}
\section{Additional Results}
\subsection{\texorpdfstring{End-to-End Evaluation at the Top-$32$ Reference Budget}{End-to-End Evaluation at the Top-32 Reference Budget}}
\label{sec:exp:top32}

Table~\ref{tab:main-e2e-top32} extends the Top-$16$ evaluation in
Table~\ref{tab:main-e2e} to a Top-$32$ reference budget, using the same five
checkpoints, full 6,500-example RULER suite, and end-to-end generation-loop
timing protocol. The baseline uses FlashMoBA with fixed Top-$32$ routing, while
\methodname{} employs dynamic threshold routing calibrated to the corresponding
average retention budget. Both variants use dense decoding with CUDA Graph
enabled, under identical evaluation settings (TP=8, batch size 1).

\methodname{} reduces end-to-end generation time across all 30 model--length
configurations, achieving per-model average speedups of 1.23--1.49$\times$.
In terms of quality, the average RULER score improves for three models, remains
essentially unchanged for Qwen3-235B-A22B, and decreases by 1.24 percentage points
for Qwen3-32B-YaRN4. These results extend the main Top-$16$ comparison and
demonstrate consistent end-to-end efficiency improvements at a larger retention
budget.

\begin{table}[htbp]
    \centering
    \caption{End-to-end RULER accuracy and generation time at the Top-$32$
    reference budget across five models and six context lengths (4K--128K).
    FlashMoBA uses fixed Top-$32$ routing, while \methodname{} uses dynamic
    threshold routing with matched average retention. Both use CUDA Graph
    decoding. Each cell reports FlashMoBA / \methodname{}; scores are percentages,
    wall times are in minutes, and speedups are shown in parentheses. The Average
    row summarizes results across the six context lengths for each model.}
    \label{tab:main-e2e-top32}
    \small
    \setlength{\tabcolsep}{2.6pt}
    \resizebox{\linewidth}{!}{\begin{tabular}{lccccc}
        \toprule
        \rowcolor{lmHeader}
        Length & Qwen3-4B & \innergqa{} & Qwen3-235B-A22B
            & Llama-3.1-8B & Qwen3-32B-YaRN4 \\
        \midrule
        \multicolumn{6}{c}{\textit{(a) RULER score: Base / Ours}} \\
        \midrule
        4K & 94.38 / 94.27 & 94.62 / 94.63 & 95.32 / 95.41 & 95.38 / 95.30 & 89.08 / 89.20 \\
        8K & 92.60 / 93.44 & 92.74 / 92.98 & 95.48 / 95.23 & 94.10 / 94.11 & 90.19 / 90.15 \\
        16K & 87.32 / 92.94 & 89.53 / 91.78 & 94.82 / 94.90 & 90.43 / 91.07 & 83.73 / 83.25 \\
        32K & 77.81 / 87.87 & 82.57 / 85.17 & 89.06 / 90.86 & 77.45 / 82.00 & 74.75 / 70.06 \\
        64K & 68.63 / 65.36 & 64.91 / 65.19 & 79.57 / 81.27 & 67.62 / 62.58 & 68.17 / 69.67 \\
        128K & 54.70 / 60.76 & 39.85 / 39.31 & 70.69 / 67.25 & 47.58 / 53.61 & 56.96 / 53.15 \\
        \midrule
        \emph{Average} & 79.24 / \textbf{82.44} & 77.37 / \textbf{78.18} & 87.49 / 87.49 & 78.76 / \textbf{79.78} & \textbf{77.15} / 75.91 \\
        \midrule
        \multicolumn{6}{c}{\textit{(b) Wall time: Base / Ours (speedup)}} \\
        \midrule
        4K & 133.4 / 70.8 (1.88$\times$) & 123.5 / 43.3 (2.85$\times$) & 323.6 / 137.6 (2.35$\times$) & 136.3 / 75.4 (1.81$\times$) & 267.8 / 171.4 (1.56$\times$) \\
        8K & 146.1 / 75.8 (1.93$\times$) & 116.3 / 42.5 (2.73$\times$) & 348.0 / 165.8 (2.10$\times$) & 144.2 / 81.6 (1.77$\times$) & 292.8 / 195.9 (1.49$\times$) \\
        16K & 169.2 / 96.4 (1.76$\times$) & 119.8 / 59.7 (2.01$\times$) & 439.3 / 237.8 (1.85$\times$) & 152.8 / 98.7 (1.55$\times$) & 343.1 / 253.2 (1.35$\times$) \\
        32K & 196.5 / 131.4 (1.49$\times$) & 165.4 / 109.9 (1.50$\times$) & 614.6 / 408.6 (1.50$\times$) & 180.6 / 134.9 (1.34$\times$) & 442.0 / 361.8 (1.22$\times$) \\
        64K & 258.0 / 192.8 (1.34$\times$) & 249.3 / 189.8 (1.31$\times$) & 972.6 / 785.6 (1.24$\times$) & 239.6 / 182.6 (1.31$\times$) & 683.7 / 589.0 (1.16$\times$) \\
        128K & 397.2 / 353.7 (1.12$\times$) & 436.9 / 369.0 (1.18$\times$) & 1739.1 / 1383.5 (1.26$\times$) & 371.9 / 323.8 (1.15$\times$) & 1270.8 / 1110.6 (1.14$\times$) \\
        \midrule
        \emph{Average} & 216.7 / \textbf{153.5} (\textbf{1.41}$\times$) & 201.8 / \textbf{135.7} (\textbf{1.49}$\times$) & 739.5 / \textbf{519.8} (\textbf{1.42}$\times$) & 204.2 / \textbf{149.5} (\textbf{1.37}$\times$) & 550.0 / \textbf{447.0} (\textbf{1.23}$\times$) \\
        \bottomrule
    \end{tabular}}
\end{table}

\FloatBarrier

\begingroup
\color{black}
\captionsetup{labelfont={color=black},textfont={color=black}}

\Needspace{9\baselineskip}
\subsection{Retention Calibration and Task-Level Diagnostics}
\label{sec:exp:main:quality}

\subsubsection{Retention Budget Calibration and Density Sweep}
\label{sec:retention-calibration}

When the matched-budget \method{} point leaves latency headroom relative to
FlashMoBA, we increase the retention budget to test whether additional retained
blocks improve quality while preserving a latency advantage. This is a budget
sweep rather than an exact latency match: each operating point is measured
independently, and we do not interpolate to force equal latency with the
baseline. Table~\ref{tab:expanded-budget} reports the threshold $\alpha$,
target retention density where applicable, generation-loop latency, and RULER
score for Qwen3-4B and \innergqa{}. Full attention serves only as a quality
reference.

\begin{table}[htbp]
    \centering
    \caption{Retention sweep and quality-latency operating points on RULER-128K.
    Each row is a separate run; $\alpha$ is the routing threshold, and
    latency is the same end-to-end generation-loop wall time in minutes as
    Table~\ref{tab:main-e2e}. The 5\%, 10\%, and 20\% rows denote the target
    retention densities used for the sweep. The full-run harness did not record
    realized routed density; the matched-budget rows instead report the fixed
    Top-16 routing's measured density. All points are measured, not interpolated.
    Full attention is a quality reference only; its latency is not matched.}
    \label{tab:expanded-budget}
    \small
    \setlength{\tabcolsep}{3pt}
        \begin{tabular}{llrrrr}
        \toprule
        \rowcolor{lmHeader}
        Model & Operating point & $\alpha$ & Target density
            & Latency (min) & RULER score \\
        \midrule
        Qwen3-4B & matched budget & 0.169 & 3.13\% & 251.6 & 46.51 \\
        Qwen3-4B & 5\% density & 0.08 & 5\% & 274.7 & 49.49 \\
        Qwen3-4B & 10\% density & 0.015 & 10\% & 346.0 & 59.45 \\
        Qwen3-4B & 20\% density & 0.003 & 20\% & 432.7 & 70.14 \\
        Qwen3-4B & full attention (reference) & \textemdash & 100\% & 475.4 & 82.85 \\
        \midrule
        \innergqa{} & matched budget & 0.173 & 3.13\% & 253.1 & 35.40 \\
        \innergqa{} & 5\% density & 0.09 & 5\% & 290.6 & 35.49 \\
        \innergqa{} & 10\% density & 0.02 & 10\% & 356.8 & 41.70 \\
        \innergqa{} & 20\% density & 0.004 & 20\% & 440.6 & 54.09 \\
        \innergqa{} & full attention (reference) & \textemdash & 100\% & 489.7 & 68.11 \\
        \bottomrule
    \end{tabular}
\end{table}

\paragraph{Retention budget verification.}
We further verify the realized retention budgets of fixed Top-16 and dynamic
threshold routing on Qwen3-4B across all six context lengths (4K--128K). The
threshold parameters are calibrated on held-out samples and fixed before
evaluation, as described in Appendix~\ref{sec:setup-details}. We then measure the
average number of retained attention tokens per query over the full RULER
evaluation, including compulsory sink and local blocks.

As shown in Table~\ref{tab:realized-budget}, dynamic threshold routing closely
matches the reference Top-16 budget across all six context lengths, with relative
deviations ranging from $-1.01\%$ to $+0.88\%$. These results confirm that the
reported comparisons operate at closely matched average retention budgets despite
the variable number of blocks selected for individual queries.

\begin{table}[htbp]
    \centering
    \caption{Verification of realized retention budgets for fixed Top-16 and dynamic
    threshold routing on Qwen3-4B across six context lengths (4K--128K). Values
    represent the average number of retained attention tokens per query over the
    full RULER evaluation, including compulsory sink and local blocks. Relative
    deviations are computed with respect to fixed Top-16 routing.}
    \label{tab:realized-budget}
    \small
    \setlength{\tabcolsep}{7pt}
    \begin{tabular}{lrrr}
        \toprule
        \rowcolor{lmHeader}
        Context length & Fixed Top-16 (tokens/query)
            & Dynamic threshold (tokens/query) & Deviation \\
        \midrule
        4K & 1{,}568 & 1{,}570 & $+0.11\%$ \\
        8K & 1{,}808 & 1{,}797 & $-0.62\%$ \\
        16K & 1{,}928 & 1{,}927 & $-0.06\%$ \\
        32K & 1{,}988 & 2{,}002 & $+0.68\%$ \\
        64K & 2{,}018 & 1{,}998 & $-1.01\%$ \\
        128K & 2{,}033 & 2{,}051 & $+0.88\%$ \\
        \bottomrule
    \end{tabular}
\end{table}

\FloatBarrier
The controlled same-backend routing comparison is reported in
Sec.~\ref{sec:exp:ablation:routing-body} (Table~\ref{tab:dynamic-vs-topk});
its quality--latency plot appears in Appendix~\ref{sec:add-figures}
(Fig.~\ref{fig:quality-latency-routing}).

\FloatBarrier
\subsubsection{Task-Level RULER Analysis}
\label{sec:task-level-ruler}

To investigate the quality differences at long context lengths, we report results
for all 13 RULER subtasks across five models at 64K and 128K under the Top-16
reference budget (Tables~\ref{tab:ruler-tasks-64k} and~\ref{tab:ruler-tasks-128k}).
These tables compare \methodname{} against the original FlashMoBA baseline,
unlike the same-backend routing comparison in Table~\ref{tab:dynamic-vs-topk}.
They provide the full task-level diagnostics for
Section~\ref{sec:exp:quality-granularity}, where we examine retention-budget and
block-granularity sensitivity in Tables~\ref{tab:ruler-budget-categories}
and~\ref{tab:block-granularity}.

\begingroup
\setlength{\intextsep}{8pt}
\setlength{\floatsep}{8pt}
\renewcommand{\arraystretch}{1.48}

\begin{table}[H]
    \centering
    \caption{Task-level RULER results at 64K under the Top-16 reference budget.
    Each cell reports Base / Ours scores (\%) with the difference
    (Ours $-$ Base) in parentheses. Average is over the 13 subtasks.
    Scores and differences are computed from unrounded values and rounded
    independently.}
    \label{tab:ruler-tasks-64k}
    \small
    \setlength{\tabcolsep}{2.6pt}
    \resizebox{\linewidth}{!}{\begin{tabular}{lccccc}
        \toprule
        \rowcolor{lmHeader}
        Subtask & Qwen3-4B & \innergqa{} & Qwen3-235B-A22B
            & Llama-3.1-8B & Qwen3-32B-YaRN4 \\
        \midrule
        NIAH Single-1 & 100.00 / 100.00 ($0.00$) & 100.00 / 100.00 ($0.00$) & 100.00 / 100.00 ($0.00$) & 99.40 / 100.00 ($+0.60$) & 100.00 / 100.00 ($0.00$) \\
        NIAH Single-2 & 79.20 / 78.80 ($-0.40$) & 99.20 / 99.00 ($-0.20$) & 100.00 / 100.00 ($0.00$) & 82.40 / 81.40 ($-1.00$) & 99.40 / 99.40 ($0.00$) \\
        NIAH Single-3 & 85.00 / 88.40 ($+3.40$) & 81.00 / 79.40 ($-1.60$) & 98.20 / 97.20 ($-1.00$) & 92.80 / 93.40 ($+0.60$) & 88.80 / 81.20 ($-7.60$) \\
        NIAH MultiKey-1 & 62.40 / 57.00 ($-5.40$) & 30.60 / 36.20 ($+5.60$) & 50.80 / 54.20 ($+3.40$) & 58.80 / 58.00 ($-0.80$) & 47.40 / 42.80 ($-4.60$) \\
        NIAH MultiKey-2 & 9.20 / 8.00 ($-1.20$) & 13.00 / 9.80 ($-3.20$) & 13.00 / 15.00 ($+2.00$) & 13.40 / 6.40 ($-7.00$) & 15.00 / 10.40 ($-4.60$) \\
        NIAH MultiKey-3 & 4.80 / 2.80 ($-2.00$) & 10.40 / 9.40 ($-1.00$) & 19.60 / 23.80 ($+4.20$) & 6.00 / 1.40 ($-4.60$) & 6.60 / 4.40 ($-2.20$) \\
        NIAH MultiQuery & 66.65 / 53.35 ($-13.30$) & 90.10 / 88.75 ($-1.35$) & 89.45 / 90.10 ($+0.65$) & 89.75 / 74.05 ($-15.70$) & 87.40 / 81.60 ($-5.80$) \\
        NIAH MultiValue & 67.30 / 66.10 ($-1.20$) & 89.95 / 89.30 ($-0.65$) & 66.90 / 69.65 ($+2.75$) & 84.05 / 86.30 ($+2.25$) & 65.10 / 58.85 ($-6.25$) \\
        VT & 95.24 / 58.28 ($-36.96$) & 55.32 / 31.56 ($-23.76$) & 97.68 / 60.36 ($-37.32$) & 84.88 / 80.64 ($-4.24$) & 88.72 / 62.16 ($-26.56$) \\
        CWE & 25.44 / 23.32 ($-2.12$) & 12.56 / 15.22 ($+2.66$) & 78.50 / 75.34 ($-3.16$) & 0.54 / 0.68 ($+0.14$) & 65.62 / 53.50 ($-12.12$) \\
        FWE & 92.87 / 86.47 ($-6.40$) & 89.80 / 93.13 ($+3.33$) & 93.80 / 95.07 ($+1.27$) & 91.87 / 90.47 ($-1.40$) & 50.00 / 62.00 ($+12.00$) \\
        QA (SQuAD) & 48.00 / 40.87 ($-7.13$) & 42.43 / 40.62 ($-1.82$) & 57.83 / 57.03 ($-0.80$) & 50.97 / 42.25 ($-8.72$) & 49.50 / 42.85 ($-6.65$) \\
        QA (HotpotQA) & 32.00 / 31.00 ($-1.00$) & 28.40 / 32.00 ($+3.60$) & 55.60 / 56.40 ($+0.80$) & 38.80 / 33.40 ($-5.40$) & 37.00 / 33.00 ($-4.00$) \\
        \midrule
        \emph{Average} & 59.08 / 53.41 ($-5.67$) & 57.14 / 55.72 ($-1.41$) & 70.87 / 68.78 ($-2.09$) & 61.05 / 57.57 ($-3.48$) & 61.58 / 56.32 ($-5.26$) \\
        \bottomrule
    \end{tabular}}
\end{table}

\begin{table}[H]
    \centering
    \caption{Task-level RULER results at 128K under the Top-16 reference budget.
    Cells follow the format of Table~\ref{tab:ruler-tasks-64k}.}
    \label{tab:ruler-tasks-128k}
    \small
    \setlength{\tabcolsep}{2.6pt}
    \resizebox{\linewidth}{!}{\begin{tabular}{lccccc}
        \toprule
        \rowcolor{lmHeader}
        Subtask & Qwen3-4B & \innergqa{} & Qwen3-235B-A22B
            & Llama-3.1-8B & Qwen3-32B-YaRN4 \\
        \midrule
        NIAH Single-1 & 100.00 / 100.00 ($0.00$) & 98.20 / 100.00 ($+1.80$) & 100.00 / 100.00 ($0.00$) & 99.60 / 98.80 ($-0.80$) & 100.00 / 100.00 ($0.00$) \\
        NIAH Single-2 & 46.20 / 52.40 ($+6.20$) & 54.80 / 60.20 ($+5.40$) & 99.00 / 98.80 ($-0.20$) & 45.20 / 40.20 ($-5.00$) & 97.20 / 98.40 ($+1.20$) \\
        NIAH Single-3 & 89.00 / 85.20 ($-3.80$) & 61.20 / 68.00 ($+6.80$) & 96.80 / 94.20 ($-2.60$) & 89.80 / 65.60 ($-24.20$) & 78.40 / 74.20 ($-4.20$) \\
        NIAH MultiKey-1 & 43.40 / 49.80 ($+6.40$) & 28.40 / 31.20 ($+2.80$) & 42.20 / 43.80 ($+1.60$) & 34.40 / 33.40 ($-1.00$) & 34.00 / 31.40 ($-2.60$) \\
        NIAH MultiKey-2 & 5.20 / 9.40 ($+4.20$) & 3.60 / 1.00 ($-2.60$) & 3.40 / 6.60 ($+3.20$) & 2.80 / 2.80 ($0.00$) & 6.40 / 4.60 ($-1.80$) \\
        NIAH MultiKey-3 & 1.40 / 6.20 ($+4.80$) & 0.60 / 0.40 ($-0.20$) & 8.20 / 12.20 ($+4.00$) & 0.60 / 0.40 ($-0.20$) & 2.00 / 2.60 ($+0.60$) \\
        NIAH MultiQuery & 46.65 / 48.95 ($+2.30$) & 24.10 / 27.55 ($+3.45$) & 88.70 / 91.65 ($+2.95$) & 34.45 / 44.95 ($+10.50$) & 83.50 / 83.40 ($-0.10$) \\
        NIAH MultiValue & 53.75 / 57.95 ($+4.20$) & 25.00 / 28.90 ($+3.90$) & 57.40 / 59.10 ($+1.70$) & 52.40 / 60.00 ($+7.60$) & 64.90 / 62.00 ($-2.90$) \\
        VT & 70.92 / 55.76 ($-15.16$) & 27.92 / 15.92 ($-12.00$) & 93.44 / 76.60 ($-16.84$) & 65.60 / 73.96 ($+8.36$) & 87.56 / 82.68 ($-4.88$) \\
        CWE & 6.34 / 5.42 ($-0.92$) & 0.56 / 0.02 ($-0.54$) & 50.76 / 51.26 ($+0.50$) & 0.30 / 0.60 ($+0.30$) & 22.56 / 17.68 ($-4.88$) \\
        FWE & 78.33 / 77.67 ($-0.67$) & 76.27 / 72.40 ($-3.87$) & 96.00 / 94.67 ($-1.33$) & 67.80 / 74.53 ($+6.73$) & 19.13 / 39.40 ($+20.27$) \\
        QA (SQuAD) & 38.20 / 35.25 ($-2.95$) & 29.32 / 27.55 ($-1.77$) & 45.98 / 44.23 ($-1.75$) & 23.15 / 22.90 ($-0.25$) & 34.38 / 33.82 ($-0.57$) \\
        QA (HotpotQA) & 23.20 / 20.60 ($-2.60$) & 24.00 / 27.00 ($+3.00$) & 47.20 / 44.60 ($-2.60$) & 22.80 / 24.60 ($+1.80$) & 31.00 / 26.80 ($-4.20$) \\
        \midrule
        \emph{Average} & 46.35 / 46.51 ($+0.15$) & 34.92 / 35.40 ($+0.48$) & 63.78 / 62.90 ($-0.87$) & 41.45 / 41.75 ($+0.30$) & 50.85 / 50.54 ($-0.31$) \\
        \bottomrule
    \end{tabular}}
\end{table}
\endgroup

The task-level results show that the quality degradation is primarily concentrated
in variable tracking (VT). At 64K, VT accuracy decreases by an average of 25.77
percentage points across the five models, accounting for approximately 55.3\% of
the reduction in the 13-task macro-average score. At 128K, the average VT
difference narrows to $-8.10$ points, indicating that the quality loss does not
consistently increase with context length. Unlike retrieval tasks that often rely
on a small number of salient evidence blocks, VT requires tracking multiple
intermediate dependencies across the context. The selection of less salient but
necessary intermediate blocks is a possible source of sensitivity to sparse
routing; the task-level measurements do not directly verify this causal mechanism.

\FloatBarrier
\endgroup

\FloatBarrier
\subsection{Comparison with Additional Sparse Attention Baselines}
\label{sec:sparse-baselines}

\begingroup
\color{black}
\captionsetup{labelfont={color=black},textfont={color=black}}

To broaden the evaluation beyond FlashMoBA, we additionally compare \methodname{}
with the original MoBA implementation and FlexPrefill \cite{lai2025flexprefill}, a representative
training-free sparse prefill method outside the MoBA family. The experiments cover
Qwen3-4B and \innergqa{} across six context lengths from 4K to 128K, using the same
6,500-example RULER evaluation framework. Table~\ref{tab:sparse-baselines} reports
the RULER scores and end-to-end generation times for all four methods.
This external-method comparison is not a controlled routing ablation: sparse
selection strategies and realized retention budgets may differ. We interpret
these results as measured quality--time operating points, separately from the
same-backend, matched-budget comparison in
Sec.~\ref{sec:exp:ablation:routing-body}.

\begin{table}[H]
    \color{black}
    \captionsetup{labelfont={color=black},textfont={color=black}}
    \centering
    \caption{Comparison with additional sparse attention baselines on the full
    6,500-example RULER suite. Results are reported for Qwen3-4B and \innergqa{}
    across six context lengths (4K--128K). (a) RULER scores (\%).
    (b) End-to-end generation-loop wall times (minutes; lower is better).
    FlashMoBA is the primary baseline used in Table~\ref{tab:main-e2e}. The methods
    employ different sparse selection strategies and may operate at different
    realized retention budgets.}
    \label{tab:sparse-baselines}
    \footnotesize
    \setlength{\tabcolsep}{8pt}
    \renewcommand{\arraystretch}{0.95}
    \begin{tabular}{llcccccc}
        \toprule
        \rowcolor{lmHeader}
        Model & Method & 4K & 8K & 16K & 32K & 64K & 128K \\
        \midrule
        \multicolumn{8}{c}{\textit{(a) RULER score (\%)}} \\
        \midrule
        Qwen3-4B & MoBA & 93.75 & 85.82 & 79.53 & 70.33 & 59.40 & 45.76 \\
        & FlashMoBA & 93.75 & 85.98 & 79.45 & 69.98 & 59.08 & 46.35 \\
        & FlexPrefill & 94.27 & 92.37 & 89.73 & 84.38 & 58.26 & 54.62 \\
        & \methodname{} & 94.29 & 91.87 & 87.18 & 79.93 & 53.41 & 46.51 \\
        \midrule
        \innergqa{} & MoBA & 94.30 & 88.47 & 82.23 & 73.79 & 58.04 & 35.47 \\
        & FlashMoBA & 94.40 & 88.20 & 82.47 & 74.09 & 57.14 & 34.92 \\
        & FlexPrefill & 94.52 & 92.74 & 82.53 & 80.47 & 65.45 & 38.64 \\
        & \methodname{} & 94.10 & 89.15 & 80.88 & 70.82 & 55.72 & 35.40 \\
        \midrule
        \multicolumn{8}{c}{\textit{(b) End-to-end time (min)}} \\
        \midrule
        Qwen3-4B & MoBA & 160.3 & 175.2 & 192.0 & 218.7 & 307.3 & 593.8 \\
        & FlashMoBA & 145.9 & 153.1 & 165.0 & 174.8 & 206.7 & 284.2 \\
        & FlexPrefill & 164.8 & 169.5 & 179.4 & 190.9 & 221.0 & 285.6 \\
        & \methodname{} & 60.5 & 69.5 & 87.7 & 114.3 & 153.1 & 251.6 \\
        \midrule
        \innergqa{} & MoBA & 139.0 & 144.0 & 143.3 & 191.6 & 297.7 & 675.4 \\
        & FlashMoBA & 119.1 & 118.7 & 116.5 & 146.2 & 178.8 & 293.0 \\
        & FlexPrefill & 143.7 & 154.5 & 155.5 & 154.2 & 194.0 & 377.2 \\
        & \methodname{} & 36.8 & 41.1 & 59.3 & 103.9 & 122.5 & 253.1 \\
        \bottomrule
    \end{tabular}
\end{table}

\paragraph{End-to-end efficiency.}
\methodname{} achieves the lowest generation time across all 12 model--length
configurations. Compared with the original MoBA implementation, it delivers
1.91--2.65$\times$ speedups on Qwen3-4B and 1.84--3.78$\times$ on \innergqa{}.
Compared with FlexPrefill, \methodname{} achieves 1.13--2.72$\times$ and
1.48--3.91$\times$ speedups on the two models, respectively. The efficiency
advantage persists at 128K, where \methodname{} reduces generation time from
285.6 to 251.6 minutes on Qwen3-4B and from 377.2 to 253.1 minutes on
\innergqa{}.

\paragraph{Quality--efficiency trade-offs.}
The compared methods exhibit different accuracy--latency operating points.
\methodname{} maintains competitive RULER accuracy relative to the original MoBA
and FlashMoBA baselines, while substantially reducing end-to-end generation time.
FlexPrefill achieves higher accuracy in most configurations, particularly at longer
context lengths, but incurs higher generation latency. For example, at 128K,
FlexPrefill achieves RULER scores of 54.62\% and 38.64\% on Qwen3-4B and
\innergqa{}, compared with 46.51\% and 35.40\% for \methodname{}. These results
highlight the distinct quality--efficiency trade-offs of the methods, while
demonstrating \methodname{}'s consistent end-to-end efficiency advantage across the
evaluated configurations.

\FloatBarrier
\endgroup

\section{Additional Setup Details}
\label{sec:setup-details}
\paragraph{Baselines}
The primary sparse baseline is MoBA's fixed Top-$k$ routing \citep{lu2025moba} executed by the FlashMoBA kernel \citep{xiao2025flashmoba}, not an unoptimized reference implementation. This compares \method{} against an optimized implementation of the same underlying block-attention method. The original-attention quality reference uses each checkpoint's native attention configuration; for hybrid models, this retains the prescribed sliding-window layers. It is a reference score, not an assumed upper bound on every task. Results identify the kernel behind each attention variant. The effect of each design component is examined through the cumulative ablation in Sec.~\ref{sec:exp:ablation:cumulative-body} (Table~\ref{tab:cumulative-implementation-ablation}), rather than inferred solely from a shared implementation codebase.

\paragraph{Training-free protocol}
For each checkpoint, model weights are held fixed across the attention variants. Neither sparse method requires additional model fine-tuning or an auxiliary learned indexer for the comparison. \method{} retains MoBA's block-scoring function and changes the block-selection rule and its execution. Training-free therefore describes the inference-time adaptation, not the training history of the checkpoint or equality of the resulting model outputs.

\paragraph{Threshold calibration}
For each model and target Top-$k$ budget, we calibrate $\alpha$ on a held-out sample that is disjoint from the test evaluation. We sweep $\alpha$ and select the value whose mean realized number of retained blocks most closely matches the corresponding Top-$k$ budget. We then verify the matched budget on a separate validation sample and fix $\alpha$ before running the reported evaluation. No task labels or benchmark scores are used during calibration.

\paragraph{Model sources}
Qwen3-4B\footnote{\url{https://huggingface.co/Qwen/Qwen3-4B-Instruct-2507}},
Qwen3-235B-A22B\footnote{\url{https://huggingface.co/Qwen/Qwen3-235B-A22B}},
Qwen3-32B-YaRN4\footnote{Qwen3-32B with YaRN-based context extension;
\url{https://huggingface.co/Qwen/Qwen3-32B}},
Llama-3.1-8B\footnote{\url{https://huggingface.co/meta-llama/Llama-3.1-8B-Instruct}},
and \innergqa{}, an internal model with no external release.

\paragraph{Metrics and measurement protocol}
We report two timing quantities and keep them separate. End-to-end results use generation-loop wall time over a complete evaluation run, a system-level measurement in which both variants execute the same dense decode path. Controlled replays instead time the complete attention path, including routing, sparse-index construction and attention computation, with component timings where available. The two are not combined into a single speedup claim, and prefill, decode and full-model inference measurements are distinguished rather than merged. Quality comparisons report task scores alongside the original-attention reference and their differences. The run specification records hardware, parallelism, precision, software and kernel versions, batch and sequence lengths, warmup and timing repetitions, the aggregation statistic, and CUDA Graph usage. Common hardware, input, precision and timing conditions are held fixed within each comparison; method-specific kernel implementations and versions are recorded separately. Throughput, memory use and kernel-utilization measurements can provide additional diagnostics where collected, but are not substitutes for the stated latency and quality comparisons. Main experiments use tensor parallelism of eight unless noted, and ablations replay on one rank.

\paragraph{Comparison protocol}
The primary comparison targets matched average retained-block budgets, with
threshold parameters calibrated and validated on held-out samples before
evaluation. Where full-evaluation retention measurements are available, we
report realized deviations explicitly
(Appendix~\ref{sec:retention-calibration}, Table~\ref{tab:realized-budget}).
Both sparse methods use the same block size and causal-boundary convention;
compulsory blocks count toward the budget. We compare latency and quality
without assuming that the selected blocks are identical. When \method{}
retains latency headroom, we additionally increase its retention budget and
report the measured quality and latency points. This supplementary sweep does not interpolate between runs or
claim an exact latency match. The original-attention configuration supplies a quality reference; latency is not matched.

\FloatBarrier
\Needspace{9\baselineskip}
\section{Additional Method Details}
\label{sec:router-details}
\subsection{Streaming Router and Sparse Metadata Construction}
\label{sec:method:nnz}

Eq.~\eqref{eq:threshold-rule} defines what is selected, but a direct implementation appears to require the full score matrix: $r_i^{\star}$ must be reduced before the scores are reused for the threshold test, and materializing $\{r_{ij}\}$ costs $\mathcal{O}(Nn)$ floating-point storage per head.

We use the two-pass streaming construction of the routing stage in Fig.~\ref{fig:overview}: the first pass computes the per-query maximum $r_i^{\star}$ over complete blocks preceding the query's own block, and the second recomputes optional-block scores against the equivalent cutoff $r_i^{\star}-\Delta$. Compulsory blocks are retained directly, emitting a candidate bitmask with per-query and per-column retained counts.

At one bit per candidate block the bitmask occupies $\mathcal{O}(Nn/32)$ machine words regardless of realized retention, and each candidate needs only a scalar comparison and a bit update, unlike an ordered Top-$k$ candidate set per query.

The execution in Sec.~\ref{sec:method:execution} consumes the transpose of
these decisions: for each KV block, the query rows that selected it. Since the
per-column counts are already available, an exclusive prefix sum determines
each column's contiguous range, after which a scatter pass writes the retained
query indices into a compact buffer. The resulting representation contains
exactly $Z$ retained query-block pairs, requires no sorting, and directly
exposes the per-column fanout used by Sec.~\ref{sec:method:fanout}.

\Needspace{12\baselineskip}
\subsection{Schedule-Length Bound}
\label{sec:method-details}

\begin{lemma}[Schedule length under list scheduling]\label{lem:makespan}
Model the $P$ resident CTA slots as identical workers and measure time in
MMA-tile service times. For column work sizes $w_{h,j}$ and
$W=\sum_{h,j}w_{h,j}$, list scheduling produces a schedule of length
$\mathcal{L}$ satisfying
\begin{equation}
    \max\Bigl(\frac{W}{P},\, \max\nolimits_{h,j} w_{h,j}\Bigr)
    \;\le\; \mathcal{L} \;\le\;
    \frac{W}{P} + \Bigl(1-\frac{1}{P}\Bigr)\max\nolimits_{h,j} w_{h,j} .
    \label{eq:makespan}
\end{equation}
\end{lemma}

\begin{algorithm}[t]
\caption{KV-Centric Ownership and Forward Pass}
\label{alg:kvcentric}
\begin{algorithmic}[1]
\Require $\mathbf{Q},\mathbf{K},\mathbf{V}\in\mathbb{R}^{N\times d}$ in HBM;
    for each head $h$ and KV block $j$, the retained row list
    $\mathcal{R}_{h,j}$ of the compact metadata of
    Sec.~\ref{sec:method:nnz}, with fanout $f_{h,j}=|\mathcal{R}_{h,j}|$;
    logical block size $B$; physical tile sizes $B_c$ and $m_{\mathrm{MMA}}$.
\State Divide $\mathbf{K},\mathbf{V}$ into $n=\lceil N/B\rceil$ logical blocks
    $\mathbf{K}_j,\mathbf{V}_j$ of size $B\times d$.
\State Allocate the partial-output workspace $\mathbf{O}_{\mathrm{tmp}}$ and the
    log-sum-exp workspace $\mathbf{L}$ in HBM.
\ForAll{$(h,j)$ \textbf{in parallel}}
    \State Load $\mathbf{K}_j,\mathbf{V}_j$ from HBM into SRAM.
        \Comment{Once per column in this conceptual formulation}
    \For{$0\le r<\lceil f_{h,j}/m_{\mathrm{MMA}}\rceil$}
        \State Gather-load the $r$-th query tile
            $\mathbf{Q}_r^{(h,j)}\gets\mathbf{Q}[\mathcal{R}_{h,j}[r
            m_{\mathrm{MMA}}:\min((r{+}1)m_{\mathrm{MMA}},f_{h,j})],h,:]$.
        \State Initialize the on-chip softmax state $(m,\ell)$ and the
            partial-output accumulator $\mathbf{O}_r$.
        \For{$0\le c<\lceil B/B_c\rceil$}
            \State $\mathbf{S}_{rc}\gets\mathbf{Q}_r^{(h,j)}
                \mathbf{K}_{j,c}^{\top}$, with causal masking.
            \State Update on-chip softmax states $(m,\ell)$ and
                partial outputs $\mathbf{O}_r$ using $\mathbf{V}_{j,c}$.
        \EndFor
        \State Store the partial output $\mathbf{O}_r$ and $m+\log\ell$ of
            these rows to $\mathbf{O}_{\mathrm{tmp}}$ and $\mathbf{L}$.
    \EndFor
\EndFor
\ForAll{query rows $i$ and heads $h$ \textbf{in parallel}}
    \State Merge all partial outputs associated with $(i,h)$ and
        write the result to $\mathbf{O}[i,h,:]$.
\EndFor
\end{algorithmic}
\end{algorithm}

\subsection{Asymmetric Data Movement}
\label{sec:method:async}

The two paths differ in kind, not merely in volume. A selected block's keys and
values occupy a contiguous window of the sequence, and the TMA moves affine
tensor regions of exactly this form \citep{nvidiacuda2024}. Routed query rows
do not form such a region: their indices come from the routing metadata and may
fall anywhere in the sequence, so they require arbitrary indexed gathers. We
accordingly move K and V with the asynchronous engine and gather query rows with
vectorized indexed loads. Within a work unit, the block's keys and values are
read by every query row the unit serves, so each CTA loads them once and keeps
them resident for its lifetime, leaving the query side as the only recurring
transfer. The matrix work follows the same asymmetry, using
$64\times64\times16$ WGMMA tiles for $QK^\top$ and $64\times128\times16$ tiles
for $PV$ \citep{nvidiaptx2024}, with the probabilities held in registers so
that the softmax between them introduces no shared-memory stage of its own.

\subsection{Pipeline Overlap and Latency Hiding}
\label{sec:method:pipeline}

\begin{figure}[H]
    \centering
    \includegraphics[width=0.88\linewidth]{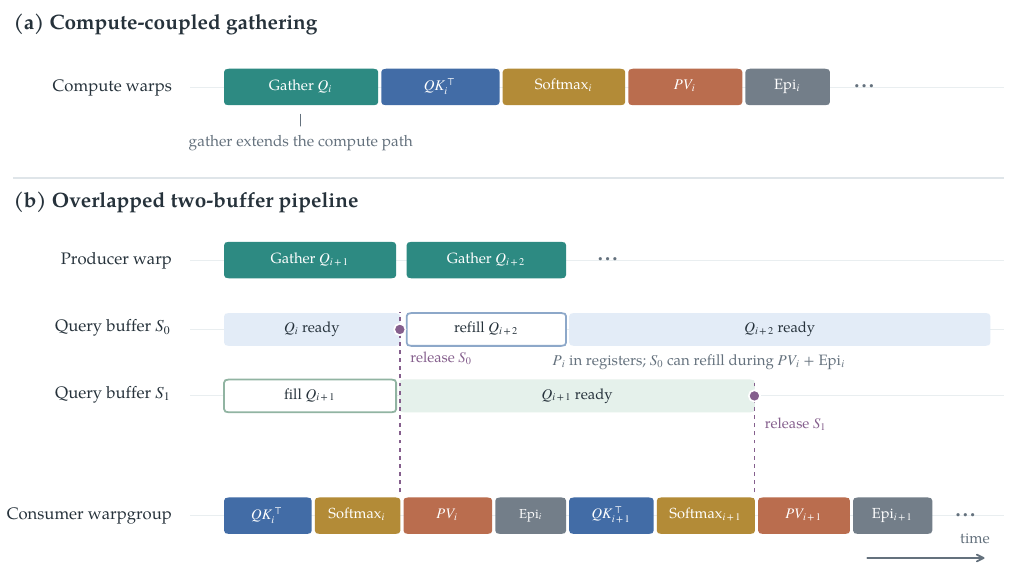}
    \caption{Two-stage query pipeline with warp specialization. (a) Without
    dedicated query-gather specialization, irregular query gathering remains
    on the compute path and serializes with matrix work. (b) A producer warp
    gathers query tile $i+1$ while the consumer warpgroup processes tile $i$;
    early slot release lets the producer refill during $PV$ and the epilogue.}
    \label{fig:pipeline}
\end{figure}

Gathering queries is the irregular half of the loop; assigning these gathers
to the compute warps interleaves indexed-load stalls with matrix instruction
issue. We therefore use one $128$-thread consumer warpgroup and one
$32$-thread producer warp. The producer decodes the next query tile's routing
metadata, including the selected query-row indices and, for each row, the
extent of its causal mask within the block, and then issues the gathers, while the consumer executes $QK^\top$, the
softmax, and $PV$. Queries flow through a two-stage ring buffer
(Fig.~\ref{fig:pipeline}), so stage $i+1$ is gathered while stage $i$ is
consumed, and index processing overlaps matrix work instead of preceding it. Once $QK^\top$ and the softmax have consumed a
staged tile we release the stage before $PV$ and the epilogue.

K/V residency removes the regular transfers from the steady-state loop, and
early release shortens a stage's shared-memory lifetime, letting the producer
refill that stage sooner. Sec.~\ref{sec:exp:ablation:pipeline} compares serial
and overlapped query gathering within the same SM90 kernel and reports the
resulting latency changes. Kernel-level stall changes are reported separately in
Appendix~\ref{sec:exp:analysis:pipeline}
(Table~\ref{tab:sm90-pipeline-ncu}). This comparison does not separately isolate
producer specialization.

\clearpage
\section{Additional Figures}
\label{sec:add-figures}
\begin{figure}[H]
    \centering
    \includegraphics[width=0.88\linewidth]{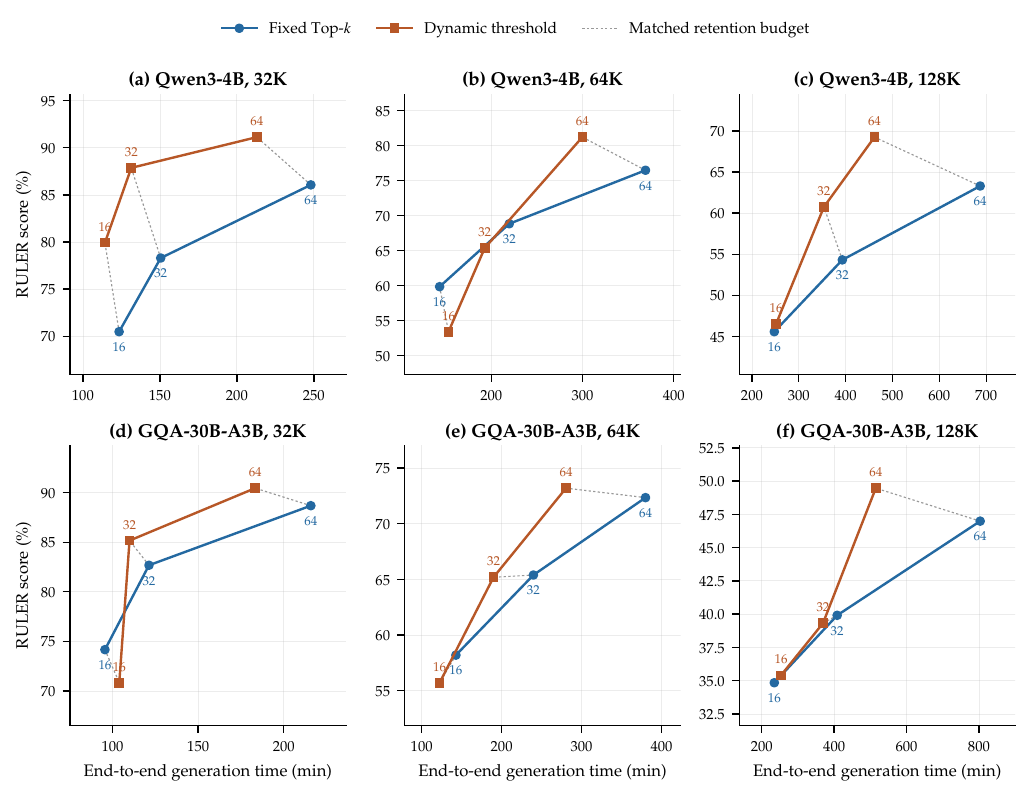}
    \caption{End-to-end RULER quality--latency trade-offs between fixed Top-$k$ and
    dynamic threshold routing using the same optimized \methodname{} execution
    backend. Each panel corresponds to one model and context length, with points
    denoting matched retention budgets of 16, 32, and 64. Lower latency and higher
    RULER scores are preferred; lines connect the measured operating points.}
    \label{fig:quality-latency-routing}
\end{figure}

\begin{figure}[H]
    \centering
    \includegraphics[width=0.48\linewidth]{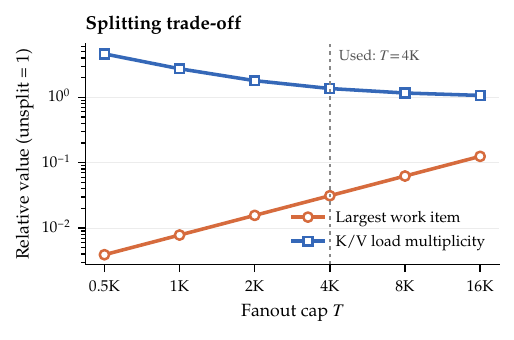}
    \caption{Fanout-aware splitting trade-off on a high-tail \innergqa{} TP8
    RULER-128K prefill trace. Relative largest work item and K/V load multiplicity
    are shown as the fanout cap $T$ varies; the dashed line marks our
    $T=4\mathrm{K}$ setting. The column-fanout distribution for this trace is
    reported in Fig.~\ref{fig:fanout-halfcol}.}
    \label{fig:fanout}
\end{figure}

\begin{figure}[H]
    \centering
    \includegraphics[width=0.82\linewidth]{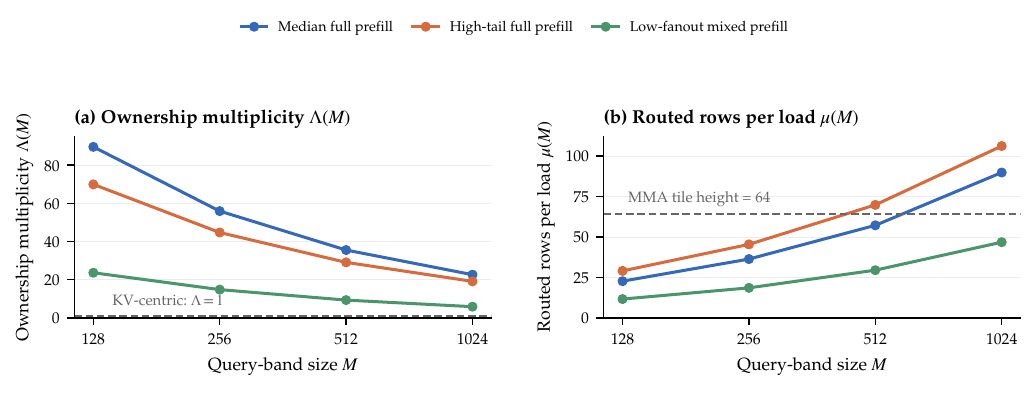}
    \caption{Ownership fragmentation under query-centric execution on three
    real Top-$16$ routing traces from TP8 RULER-128K inference with
    \innergqa{}. (a) Mean ownership multiplicity; KV-centric ownership gives
    $\Lambda=1$. (b) Mean routed rows per load; the dashed line marks the
    64-row MMA tile height.}
    \label{fig:ownership}
\end{figure}

\clearpage
\Needspace*{12\baselineskip}
\section{Proofs}
\label{sec:proofs}

\paragraph{Notation.}
Let $\beta_M(i)=\lfloor i/M\rfloor$ be the index of the band owning query row
$i$, so band $b$ is $\mathcal{B}_b(M)=\{bM,\ldots,(b{+}1)M-1\}$. For a selected
column $(h,j)$ write
\[
    i^{\min}_{h,j}=\min\mathcal{Q}_{h,j},\qquad
    i^{\max}_{h,j}=\max\mathcal{Q}_{h,j},\qquad
    s_{h,j}=i^{\max}_{h,j}-i^{\min}_{h,j}+1 .
\]
Under this notation Eq.~\eqref{eq:ownership} becomes
\[
    \lambda_{h,j}(M)=\bigl|\beta_M(\mathcal{Q}_{h,j})\bigr|,\qquad
    \Lambda(M)=\frac{1}{C}\sum_{h,j}\lambda_{h,j}(M).
\]

\begin{proof}[Proof of Lemma~\ref{lem:span}]
Write $\lambda_{h,j}=\lambda_{h,j}(M)$. Since every selected column has
$\mathcal{Q}_{h,j}\neq\emptyset$, we have $\beta_M(\mathcal{Q}_{h,j})\neq\emptyset$ and hence
\[
    \lambda_{h,j}\ \ge\ 1 \qquad\text{for every selected }(h,j).
\]
Assume $\Lambda(M)=1$, i.e.
\[
    \sum_{h,j}\lambda_{h,j}=C.
\]
Because each $\lambda_{h,j}$ is a positive integer and there are exactly $C$
terms, the two displays force
\[
    \lambda_{h,j}=1 \qquad\text{for every selected }(h,j).
\]
Fix $(h,j)$. Then $\bigl|\beta_M(\mathcal{Q}_{h,j})\bigr|=1$ means that
$\mathcal{Q}_{h,j}\subseteq\mathcal{B}_b(M)$ for a single $b$, hence
\[
    bM\ \le\ i^{\min}_{h,j}\ \le\ i^{\max}_{h,j}\ \le\ (b{+}1)M-1 .
\]
Consequently
\[
    s_{h,j}
    = i^{\max}_{h,j}-i^{\min}_{h,j}+1
    \ \le\ \bigl((b{+}1)M-1\bigr)-bM+1
    \ =\ M .
\]
Taking the maximum over all selected columns gives
\[
    M\ \ge\ \max_{h,j}s_{h,j},
\]
which is Eq.~\eqref{eq:span-bound}.

The implication cannot be reversed. Take $M=128$ and a routing pattern with a
single selected column ($C=1$) and $\mathcal{Q}=\{127,128\}$. Then
\[
    s = 128-127+1 = 2\ \le\ 128 = M,
\]
whereas $\beta_M(127)=0$ and $\beta_M(128)=1$, so
$\lambda=|\{0,1\}|=2$ and $\Lambda(M)=2\neq1$. Hence $M\ge\max_{h,j}s_{h,j}$ is
necessary but not sufficient for $\Lambda(M)=1$.
\end{proof}

\begin{proof}[Proof of Lemma~\ref{lem:makespan}]
The classical list-scheduling bound of \citet{graham1966} gives the result;
we restate its short proof for completeness. Index the work units by
$\ell=1,\ldots,n$ with sizes $w_{\ell}\ge0$, and write
\[
    W=\sum_{\ell=1}^{n}w_{\ell},\qquad
    w_{\max}=\max_{1\le\ell\le n}w_{\ell}.
\]
All units are available at time $0$, run non-preemptively on $P$ identical
workers, and list scheduling starts an available unit whenever a worker is idle.

\emph{Lower bound.} The unit of size $w_{\max}$ occupies one worker for
$w_{\max}$ time, so $\mathcal{L}\ge w_{\max}$; and $P$ workers can complete at
most $P\mathcal{L}$ work by time $\mathcal{L}$, so $W\le P\mathcal{L}$. Together,
\[
    \mathcal{L}\ \ge\ \max\Bigl\{\frac{W}{P},\,w_{\max}\Bigr\}.
\]

\emph{Upper bound.} Let $J$ be a unit that completes last, with start time
$s_J$ and size $w_J$, so that
\[
    \mathcal{L}=s_J+w_J .
\]
For every $t\in[0,s_J)$ no worker is idle: if some worker were idle at such a
$t$, then $J$ would still be unstarted and available, and list scheduling would
have started it at $t$, contradicting the definition of $s_J$. Hence all $P$
workers execute work throughout $[0,s_J)$, while $J$ itself is processed only
after $s_J$. The corresponding work inequality is
\[
    P\,s_J\ \le\ W-w_J
    \qquad\Longrightarrow\qquad
    s_J\ \le\ \frac{W-w_J}{P}.
\]
Substituting this into the expression for $\mathcal{L}$,
\[
    \mathcal{L}
    = s_J+w_J
    \ \le\ \frac{W-w_J}{P}+w_J
    \ =\ \frac{W}{P}+\Bigl(1-\frac{1}{P}\Bigr)w_J
    \ \le\ \frac{W}{P}+\Bigl(1-\frac{1}{P}\Bigr)w_{\max},
\]
where the last step uses $w_J\le w_{\max}$. Identifying $\ell$ with $(h,j)$ and
substituting $w_{h,j}=\lceil f_{h,j}/m_{\mathrm{MMA}}\rceil$ from
Eq.~\eqref{eq:worksize} yields Eq.~\eqref{eq:makespan}.
\end{proof}


\begin{thebibliography}{33}
\providecommand{\natexlab}[1]{#1}
\providecommand{\url}[1]{\texttt{#1}}
\expandafter\ifx\csname urlstyle\endcsname\relax
  \providecommand{\doi}[1]{doi: #1}\else
  \providecommand{\doi}{doi: \begingroup \urlstyle{rm}\Url}\fi

\bibitem[Lu et~al.(2025)Lu, Jiang, Liu, Du, Jiang, Hong, Liu, He, Yuan, Wang,
  Huang, Yuan, Xu, Xu, Lai, Chen, Zheng, Yan, Su, Wu, Zhang, Yang, Zhou, Zhang,
  and Qiu]{lu2025moba}
Enzhe Lu, Zhejun Jiang, Jingyuan Liu, Yulun Du, Tao Jiang, Chao Hong, Shaowei
  Liu, Weiran He, Enming Yuan, Yuzhi Wang, Zhiqi Huang, Huan Yuan, Suting Xu,
  Xinran Xu, Guokun Lai, Yanru Chen, Huabin Zheng, Junjie Yan, Jianlin Su,
  Yuxin Wu, Yutao Zhang, Zhilin Yang, Xinyu Zhou, Mingxing Zhang, and Jiezhong
  Qiu.
\newblock {MoBA}: Mixture of block attention for long-context {LLMs}.
\newblock In \emph{Advances in Neural Information Processing Systems},
  volume~38, 2025.

\bibitem[Xiao et~al.(2025)Xiao, Guo, Mazaheri, and Han]{xiao2025flashmoba}
Guangxuan Xiao, Junxian Guo, Kasra Mazaheri, and Song Han.
\newblock Optimizing mixture of block attention.
\newblock \emph{arXiv preprint arXiv:2511.11571}, 2025.

\bibitem[Beltagy et~al.(2020)Beltagy, Peters, and Cohan]{beltagy2020longformer}
Iz~Beltagy, Matthew~E. Peters, and Arman Cohan.
\newblock Longformer: The long-document transformer.
\newblock \emph{arXiv preprint arXiv:2004.05150}, 2020.

\bibitem[Zaheer et~al.(2020)Zaheer, Guruganesh, Dubey, Ainslie, Alberti,
  Ontanon, Pham, Ravula, Wang, Yang, and Ahmed]{zaheer2020bigbird}
Manzil Zaheer, Guru Guruganesh, Kumar~Avinava Dubey, Joshua Ainslie, Chris
  Alberti, Santiago Ontanon, Philip Pham, Anirudh Ravula, Qifan Wang, Li~Yang,
  and Amr Ahmed.
\newblock {Big Bird}: Transformers for longer sequences.
\newblock In \emph{Advances in Neural Information Processing Systems},
  volume~33, 2020.

\bibitem[Kitaev et~al.(2020)Kitaev, Kaiser, and Levskaya]{kitaev2020reformer}
Nikita Kitaev, {\L}ukasz Kaiser, and Anselm Levskaya.
\newblock Reformer: The efficient transformer.
\newblock In \emph{International Conference on Learning Representations}, 2020.

\bibitem[Roy et~al.(2021)Roy, Saffar, Vaswani, and Grangier]{roy2021routing}
Aurko Roy, Mohammad Saffar, Ashish Vaswani, and David Grangier.
\newblock Efficient content-based sparse attention with routing transformers.
\newblock \emph{Transactions of the Association for Computational Linguistics},
  9:\penalty0 53--68, 2021.

\bibitem[Shen et~al.(2025)Shen, Lu, Gui, Li, He, Yin, and Sun]{shen2025ssa}
Zhenyi Shen, Junru Lu, Lin Gui, Jiazheng Li, Yulan He, Di~Yin, and Xing Sun.
\newblock {SSA}: Sparse sparse attention by aligning full and sparse attention
  outputs in feature space.
\newblock \emph{arXiv preprint arXiv:2511.20102}, 2025.

\bibitem[Yuan et~al.(2025{\natexlab{a}})Yuan, Gao, Dai, Luo, Zhao, Zhang, Xie,
  Wei, Wang, Xiao, Wang, Ruan, Zhang, Liang, and Zeng]{yuan2025nsa}
Jingyang Yuan, Huazuo Gao, Damai Dai, Junyu Luo, Liang Zhao, Zhengyan Zhang,
  Zhenda Xie, Y.~X. Wei, Lean Wang, Zhiping Xiao, Yuqing Wang, Chong Ruan, Ming
  Zhang, Wenfeng Liang, and Wangding Zeng.
\newblock Native sparse attention: Hardware-aligned and natively trainable
  sparse attention.
\newblock In \emph{Annual Meeting of the Association for Computational
  Linguistics}, 2025{\natexlab{a}}.

\bibitem[{DeepSeek-AI}(2025)]{deepseekai2025dsa}
{DeepSeek-AI}.
\newblock {DeepSeek-V3.2}: Pushing the frontier of open large language models.
\newblock \emph{arXiv preprint arXiv:2512.02556}, 2025.

\bibitem[Lai et~al.(2026)Lai, Xu, Yang, Chen, Xu, Zeng, Li, Sun, Zhu, Zhang,
  Hu, Li, Gao, Li, Zhu, Zhou, and Zhao]{lai2026msa}
Xunhao Lai, Weiqi Xu, Yufeng Yang, Qiaorui Chen, Yang Xu, Lunbin Zeng, Xiaolong
  Li, Haohai Sun, Haichao Zhu, Vito Zhang, Jinkai Hu, Jiayao Li, Rui Gao, Zekun
  Li, Songquan Zhu, Jingkai Zhou, and Pengyu Zhao.
\newblock Minimax sparse attention.
\newblock \emph{arXiv preprint arXiv:2606.13392}, 2026.

\bibitem[Hu et~al.(2026)Hu, Wei, Gu, Zhang, Liang, Li, Zhu, Wang, Han, Bai, Tu,
  Mi, and Liang]{hu2026hils}
Xiang Hu, Xinyu Wei, Hao Gu, Minshen Zhang, Tian Liang, Huayang Li, Lei Zhu,
  Yan Wang, Sirui Han, Yushi Bai, Kewei Tu, Haitao Mi, and Leo Liang.
\newblock Hierarchical sparse attention done right: Toward infinite context
  modeling.
\newblock \emph{arXiv preprint arXiv:2607.02980}, 2026.

\bibitem[Gao et~al.(2026)Gao, Wei, Zhang, Cheng, Chen, Tang, Jiang, Song,
  Zhang, Zhao, Yang, Wang, Cao, and Luo]{gao2026hysparse}
Yizhao Gao, Jianyu Wei, Qihao Zhang, Yu~Cheng, Shimao Chen, Zhengju Tang, Zihan
  Jiang, Yifan Song, Hailin Zhang, Liang Zhao, Bo~Yang, Gang Wang, Shijie Cao,
  and Fuli Luo.
\newblock {HySparse}: A hybrid sparse attention architecture with oracle token
  selection and {KV} cache sharing.
\newblock \emph{arXiv preprint arXiv:2602.03560}, 2026.

\bibitem[Wei et~al.(2026)Wei, Gao, Zhang, Chen, Tang, Cheng, Zhou, Jiang, Song,
  Zhang, Zhao, Yang, Wang, Cao, and Luo]{wei2026hysparse2}
Jianyu Wei, Yizhao Gao, Qihao Zhang, Shimao Chen, Zhengju Tang, Yu~Cheng,
  Shengjie Zhou, Zihan Jiang, Yifan Song, Hailin Zhang, Liang Zhao, Bo~Yang,
  Gang Wang, Shijie Cao, and Fuli Luo.
\newblock {HySparse2}: Hybrid sparse attention with two-level {KV} sharing.
\newblock \emph{arXiv preprint arXiv:2609.26368}, 2026.

\bibitem[Qiu et~al.(2026)Qiu, Wang, Li, Li, Xu, Wang, Zhang, Men, Mao, Zhang,
  Zhou, Luo, Huang, Lian, Huang, Chen, Zhang, Xu, Wang, Chen, Wang, Jiang,
  Yuan, Sun, Jin, Zhang, Wang, Ren, Wang, Zhang, Dong, Cao, Ma, Mao, Zheng, and
  Liu]{qiu2026qsa}
Zihan Qiu, Zekun Wang, Xiao Li, Yanpeng Li, Yang Xu, Yixuan Wang, Huaqing
  Zhang, Rui Men, Bochao Mao, Chengruidong Zhang, Fan Zhou, Hao Luo, Haofeng
  Huang, Haoran Lian, Haoyan Huang, Hongqing Chen, Jianwei Zhang, Jing Xu,
  Junjie Wang, Langshi Chen, Liangyu Wang, Linlang Jiang, Man Yuan, Minmin Sun,
  Peng Jin, Siqi Zhang, Siyu Wang, Xingzhang Ren, Yakai Wang, Yi~Zhang, Yiming
  Dong, Yizhong Cao, Yubo Ma, Yunfei Mao, Bo~Zheng, and Dayiheng Liu.
\newblock On the design of {Qwen3.8-Next} architecture: Evaluation, efficiency,
  and training stability.
\newblock \emph{arXiv preprint arXiv:2608.30320}, 2026.

\bibitem[Liu et~al.(2026)Liu, Cheng, Niu, Su, Xue, Liu, Yu, and
  Zhu]{liu2026pivot}
Hong Liu, Yuan Cheng, Lin Niu, Yi~Su, Yufei Xue, Anmin Liu, Guanghua Yu, and
  Jianchen Zhu.
\newblock {PIVOT}: Efficient query-group indexing for token-level sparse
  attention.
\newblock \emph{arXiv preprint arXiv:2607.24593}, 2026.

\bibitem[Bai et~al.(2026)Bai, Dong, Jiang, Lv, Du, Zeng, Tang, and
  Li]{bai2026indexcache}
Yushi Bai, Qian Dong, Ting Jiang, Xin Lv, Zhengxiao Du, Aohan Zeng, Jie Tang,
  and Juanzi Li.
\newblock {IndexCache}: Accelerating sparse attention via cross-layer index
  reuse.
\newblock \emph{arXiv preprint arXiv:2603.12201}, 2026.

\bibitem[{DeepSeek-AI}(2026)]{deepseekai2026csa2}
{DeepSeek-AI}.
\newblock {DeepSeek-V4.1-Flash}: Pushing the limits of {KV} cache compression.
\newblock \emph{arXiv preprint arXiv:2609.19969}, 2026.

\bibitem[Yuan et~al.(2025{\natexlab{b}})Yuan, Shinn, Xu, Cui, Klimiashvili,
  Xiao, Zheng, Li, Zhou, Ye, You, Zheng, Brown, Wang, Hoehnerbach, Cai,
  Demouth, Owens, Hu, Han, Liu, and Mao]{yuan2025blasst}
Jiayi Yuan, Cameron Shinn, Kai Xu, Jingze Cui, George Klimiashvili, Guangxuan
  Xiao, Perkz Zheng, Bo~Li, Yuxin Zhou, Zhouhai Ye, Weijie You, Tian Zheng,
  Dominic Brown, Pengbo Wang, Markus Hoehnerbach, Richard Cai, Julien Demouth,
  John~D. Owens, Xia Hu, Song Han, Timmy Liu, and Huizi Mao.
\newblock {BLASST}: Dynamic {BLocked} attention sparsity via softmax
  thresholding.
\newblock \emph{arXiv preprint arXiv:2512.12087}, 2025{\natexlab{b}}.

\bibitem[Wu et~al.(2026)Wu, Hou, Yang, Tian, Wan, Zhang, and Tong]{wu2026vmoba}
Jianzong Wu, Liang Hou, Haotian Yang, Ye~Tian, Pengfei Wan, Di~Zhang, and
  Yunhai Tong.
\newblock {VMoBA}: Mixture-of-block attention for video diffusion models.
\newblock In \emph{International Conference on Learning Representations}, 2026.

\bibitem[Fan et~al.(2026)Fan, Huang, Wu, Wang, Wang, and
  He]{fan2026flashprefill}
Qihang Fan, Huaibo Huang, Zhiying Wu, Juqiu Wang, Bingning Wang, and Ran He.
\newblock {FlashPrefill}: Instantaneous pattern discovery and thresholding for
  ultra-fast long-context prefilling.
\newblock \emph{arXiv preprint arXiv:2603.06199}, 2026.

\bibitem[Lai et~al.(2025)Lai, Lu, Luo, Ma, and Zhou]{lai2025flexprefill}
Xunhao Lai, Jianqiao Lu, Yao Luo, Yiyuan Ma, and Xun Zhou.
\newblock {FlexPrefill}: A context-aware sparse attention mechanism for
  efficient long-sequence inference.
\newblock In \emph{The Thirteenth International Conference on Learning
  Representations}, 2025.
\newblock URL \url{https://openreview.net/forum?id=OfjIlbelrT}.

\bibitem[Zhou et~al.(2026)Zhou, Li, Tang, Li, Liu, Lan, Qu, Yao, and
  Ma]{zhou2026rtpurbo}
Yanke Zhou, Yiduo Li, Hanlin Tang, Maohua Li, Kan Liu, Tao Lan, Lin Qu, Yuan
  Yao, and Xiaoxing Ma.
\newblock Full attention strikes back: Transferring full attention into sparse
  within hundred training steps.
\newblock \emph{arXiv preprint arXiv:2605.16928}, 2026.

\bibitem[Ahmad and Yun(2026)]{ahmad2026spotattention}
Huzama Ahmad and Se-Young Yun.
\newblock {SpotAttention}: Plug-in block-sparse routing for pretrained
  long-context transformers.
\newblock \emph{arXiv preprint arXiv:2606.22874}, 2026.

\bibitem[Jayanth and Prasanna(2026)]{jayanth2026fastprefill}
Rakshith Jayanth and Viktor~K. Prasanna.
\newblock {FAST-Prefill}: {FPGA} accelerated sparse attention for long context
  {LLM} prefill.
\newblock In \emph{2026 IEEE 34th Annual International Symposium on
  Field-Programmable Custom Computing Machines (FCCM)}, 2026.
\newblock \doi{10.1109/FCCM68464.2026.00067}.

\bibitem[Yan et~al.(2025)Yan, Jiang, Chen, Mai, Chen, and Yuan]{yan2025fsa}
Ran Yan, Youhe Jiang, Zhuoming Chen, Haohui Mai, Beidi Chen, and Binhang Yuan.
\newblock {FSA}: An alternative efficient implementation of native sparse
  attention kernel.
\newblock \emph{arXiv preprint arXiv:2508.18224}, 2025.

\bibitem[Dao et~al.(2022)Dao, Fu, Ermon, Rudra, and
  R{\'e}]{dao2022flashattention}
Tri Dao, Daniel~Y. Fu, Stefano Ermon, Atri Rudra, and Christopher R{\'e}.
\newblock {FlashAttention}: Fast and memory-efficient exact attention with
  {IO}-awareness.
\newblock In \emph{Advances in Neural Information Processing Systems},
  volume~35, 2022.

\bibitem[Dao(2024)]{dao2023flashattention2}
Tri Dao.
\newblock {FlashAttention-2}: Faster attention with better parallelism and work
  partitioning.
\newblock In \emph{The Twelfth International Conference on Learning
  Representations}, 2024.

\bibitem[Shah et~al.(2024)Shah, Bikshandi, Zhang, Thakkar, Ramani, and
  Dao]{shah2024flashattention3}
Jay Shah, Ganesh Bikshandi, Ying Zhang, Vijay Thakkar, Pradeep Ramani, and Tri
  Dao.
\newblock {FlashAttention-3}: Fast and accurate attention with asynchrony and
  low-precision.
\newblock In \emph{Advances in Neural Information Processing Systems},
  volume~37, 2024.

\bibitem[Zadouri et~al.(2026)Zadouri, Hoehnerbach, Shah, Liu, Thakkar, and
  Dao]{zadouri2026flashattention4}
Ted Zadouri, Markus Hoehnerbach, Jay Shah, Timmy Liu, Vijay Thakkar, and Tri
  Dao.
\newblock {FlashAttention-4}: Algorithm and kernel pipelining co-design for
  asymmetric hardware scaling.
\newblock \emph{arXiv preprint arXiv:2603.05451}, 2026.

\bibitem[Kwon et~al.(2023)Kwon, Li, Zhuang, Sheng, Zheng, Yu, Gonzalez, Zhang,
  and Stoica]{kwon2023pagedattention}
Woosuk Kwon, Zhuohan Li, Siyuan Zhuang, Ying Sheng, Lianmin Zheng, Cody~Hao Yu,
  Joseph~E. Gonzalez, Hao Zhang, and Ion Stoica.
\newblock Efficient memory management for large language model serving with
  {PagedAttention}.
\newblock In \emph{Proceedings of the ACM SIGOPS 29th Symposium on Operating
  Systems Principles}, 2023.

\bibitem[{NVIDIA}(2024{\natexlab{a}})]{nvidiacuda2024}
{NVIDIA}.
\newblock {CUDA} {P}rogramming {G}uide, {V}ersion 12.4.
\newblock Technical report, {NVIDIA} Corporation, 2024{\natexlab{a}}.
\newblock [Online]. Available:
  \url{https://docs.nvidia.com/cuda/cuda-c-programming-guide/index.html}.

\bibitem[{NVIDIA}(2024{\natexlab{b}})]{nvidiaptx2024}
{NVIDIA}.
\newblock {P}arallel {T}hread {E}xecution {ISA}, {V}ersion 8.4.
\newblock Technical report, {NVIDIA} Corporation, 2024{\natexlab{b}}.
\newblock [Online]. Available:
  \url{https://docs.nvidia.com/cuda/pdf/ptx_isa_8.4.pdf}.

\bibitem[Graham(1966)]{graham1966}
Ronald~L. Graham.
\newblock Bounds for certain multiprocessing anomalies.
\newblock \emph{Bell System Technical Journal}, 45\penalty0 (9):\penalty0
  1563--1581, 1966.
\newblock \doi{10.1002/j.1538-7305.1966.tb01709.x}.

\end{thebibliography}
\end{document}